\documentclass[10pt]{article}
\usepackage{multicol}
\usepackage{graphicx}
\usepackage{amsmath}
\usepackage[a4paper]{geometry}
\usepackage{rotating}
\usepackage{verbatim}
\usepackage{hyperref}

\begin{document}

\begin{center}
{\Large\bf Excitation Functions of Tsallis-like Parameters in
High-Energy Nucleus-Nucleus Collisions}

\vskip.75cm

Li-Li Li$^{1}$, Fu-Hu Liu$^{1,*}$ and Khusniddin K. Olimov$^{2,*}$

\vskip.25cm

\small{$^1$ Institute of Theoretical Physics \& Collaborative
Innovation Center of Extreme Optics \& State Key Laboratory of
Quantum Optics and Quantum Optics Devices, Shanxi University,
Taiyuan 030006, China; shanxi\_lll@163.com (L.L.L.)

$^2$ Laboratory of High Energy Physics, Physical-Technical
Institute of SPA ``Physics-Sun" of Uzbek Academy of Sciences,
Chingiz Aytmatov str. $2^b$, 100084 Tashkent, Uzbekistan

$^*$ Correspondence: fuhuliu@163.com or fuhuliu@sxu.edu.cn
(F.H.L.); khkolimov@gmail.com (K.K.O.)}

\end{center}

\vskip.5cm

{\bf Abstract:} The transverse momentum spectra of charged pions,
kaons, and protons produced at mid-rapidity in central
nucleus-nucleus (AA) collisions at high energies are analyzed by
considering particles to be created from two participant partons
which are assumed to be contributors from the collision system.
Each participant (contributor) parton is assumed to contribute to
the transverse momentum by a Tsallis-like function. The
contributions of the two participant partons are regarded as the
two components of transverse momentum of the identified particle.
The experimental data measured in high-energy AA collisions by
international collaborations are studied. The excitation functions
of kinetic freeze-out temperature and transverse flow velocity are
extracted. The two parameters increase quickly from $\approx3$ to
$\approx10$ GeV (exactly from 2.7 to 7.7 GeV) and then slowly at
above 10 GeV with the increase of collision energy. In particular,
there is a plateau from near 10 GeV to 200 GeV in the excitation
function of kinetic freeze-out temperature.
\\

\\


\vskip1.5cm

\begin{multicols}{2}

\section{Introduction}

High-energy collider experiments are designed to study the
strongly interacting matter at high temperatures and
densities~\cite{7}. The deconfinement of colliding hadrons into
quark-gluon plasma (QGP), which then rapidly expands and cools
down~\cite{8}, is conjectured to be created at such extreme
collision energies~\cite{9,10,11,12}. In high energy and nuclear
physics, the study of transverse [momentum (\(p_T\)) or mass
($m_T$)] spectra of charged particles produced in nucleus-nucleus
(AA) collisions is very important. In particular, the AA collision
process at the Relativistic Heavy Ion Collider (RHIC) and the
Large Hadron Collider (LHC) provides a good opportunity to study
the signals and characteristics of QGP generation, so as to
indirectly study the system evolution and the reaction mechanism
of particle generation.

During the time evolution of collision system~\cite{13,14,15}, the
stages of kinetic freeze-out and chemical freeze-out are two
important processes. In the stage of chemical freeze-out, a phase
transition from QGP to hadrons occurred in the system, so the
composition and ratio of various particles remain unchanged. In
the stage of kinetic freeze-out, elastic collisions among
particles stop, so their $p_T$ and then $m_T$ spectra are
unchanged~\cite{14,16}. Therefore, by studying the $p_T$ ($m_T$)
spectra, we can obtain some useful information, such as the
effective temperature ($T$), the chemical freeze-out temperature
($T_{ch}$), and the kinetic freeze-out temperature ($T_0$ or
$T_{kin}$) of the system, as well as the transverse flow velocity
($\beta_T$) of the final state particles. The temperature in which
we do not exclude the contribution of transverse flow is called
the effective temperature which is related to the kinetic
freeze-out temperature. The temperatures in the stages of chemical
and kinetic freeze-outs are called the chemical and kinetic
freeze-out temperatures respectively.

It is very important to study the behavior of $T_0$ and $\beta_T$
due to their relation to map the phase diagram of Quantum
Chromodynamics (QCD), though $T_{ch}$ is usually
used~\cite{43,44,45,46,47,48} in the phase diagram. In order to
extract $T_0$ and $\beta_T$, and study their dependence on energy,
we can analyze the $p_T$ ($m_T$) spectra of particles using
different models. These models include, but are not limited to,
the blast-wave model with Boltzmann-Gibbs statistics~\cite{17,18}
or Tsallis statistics~\cite{19,20,21}, as well as other
alternative methods~\cite{22,23,24,25,26} based on the standard
distribution or Tsallis distribution. Here, the standard
distribution denotes together the Boltzmann, Fermi-Dirac, and
Bose-Einstein distributions. The alternative method regards the
intercept of $T$ versus $m_0$ as $T_0$, and the slope of $\langle
p_T\rangle$ versus $\overline m$ as $\beta_T$, where $m_0$,
$\langle p_T\rangle$, and $\overline m$ denote the rest mass, mean
$p_T$, and mean energy of the given particles, respectively.

In our recent work~\cite{28,29}, the blast-wave model with
Boltzmann-Gibbs statistics or Tsallis statistics and the standard
distribution have been used to analyze the spectra of particles
produced in high-energy proton-proton (pp) and AA collisions. The
related parameters were extracted and their excitation functions
were obtained. Not only the blast-wave model~\cite{17,18,19,20,21}
but also the alternative method~\cite{22,23,24,25,26} can be used
to extract $T_0$ and $\beta_T$, though an effective temperature
$T$ is used in the latter. The alternative method is partly a new
one, in which the extractions of both $T_0$ and $\beta_T$ are
based on $T$~\cite{22,23,30} and the related derived quantities
such as $\langle p_T\rangle$ and $\overline m$.

Due to the importance of $T_0$ and $\beta_T$ and their excitation
functions, we use a new method in the framework of multisource
thermal model~\cite{32a} to describe the $p_T$ ($m_T$) spectra of
identified particles in this work. Considering the contributions
of two participant (contributor) partons to $p_T$ of a given
particle, we regard the two contributions as the two components of
$p_T$. The $p_T$ ($m_T$) spectra of identified particles
(concretely charged pions, kaons, and protons) produced at
mid-rapidity (mid-$y$) in central AA collisions which include
gold-gold (Au-Au) collisions at the Alternating Gradient
Synchrotron (AGS), lead-lead (Pb-Pb) collisions at the Super
Proton Synchrotron (SPS), Au-Au collisions at the RHIC, and Pb-Pb
and xenon-xenon (Xe-Xe) collisions at the LHC are studied. The
center-of-mass energy per nucleon pair, $\sqrt{s_{NN}}$,
considered by us is from 2.7 GeV to 5.44 TeV. After fitting the
experimental data measured by the E866~\cite{E866},
E895~\cite{E8951,E8952}, E802~\cite{E802,E8022},
NA49~\cite{NA49,NA492}, STAR~\cite{1,2,3}, and ALICE
Collaborations~\cite{4,5,6}, we analyze the tendency of
parameters.

The remainder of this paper is structured as follows. The
formalism and method are shortly described in Section 2. Results
and discussion are given in Section 3. In Section 4, we summarize
our main observations and conclusions.
\\

\section{Formalism and method}

The Tsallis distribution has different forms or
revisions~\cite{36,36a,36aa,36aaa}, we have the Tsallis-like
distribution of $p_T$ at mid-$y$ to be
\begin{align}
\frac{d^2N}{dydp_T} \propto \frac{dN}{dy} m_T
\bigg[1+\frac{(q-1)(m_T-\mu-m_0)}{T}\bigg]^{-1/(q-1)},
\end{align}
where $N$ denotes the number of particles,
\begin{align}
m_T=\sqrt{p^2_T+m^2_0}
\end{align}
can be obtained using $p_T$,
\begin{align}
q=1+\frac{1}{n}
\end{align}
is an entropy index that characterizes the degree of equilibrium
or non-equilibrium, $n$ is a parameter related to $q$, and $\mu$
is the chemical potential. In particular, in the expression of
$m_T-\mu-m_0$, $m_T$ is simplified from $m_T\cosh y$ because
$\cosh y\approx1$ at mid-$y$.

We have the probability density function of $p_T$ at mid-$y$ to be
\begin{align}
\frac{1}{N}\frac{dN}{dp_T} \propto m_T
\bigg[1+\frac{(q-1)(m_T-\mu-m_0)}{T}\bigg]^{-1/(q-1)}.
\end{align}
Empirically, to fit the spectra of $p_T$ at mid-$y$ in this work,
Eq. (4) can be revised as
\begin{align}
f(p_T,T) &= C m_T^{a_0}
\bigg[1+\frac{(q-1)(m_T-\mu-m_0)}{T}\bigg]^{-1/(q-1)},
\end{align}
where $C$ is the normalization constant, $a_0$ is a new
non-dimensional parameter that describes the bending degree of the
distribution in low-$p_T$ region ($p_T=0\sim1$ GeV/$c$), which is
introduced artificially and tested in our recent
work~\cite{36b,36bb}, and $m_T^{a_0}$ is revised from $m_T$ due to
the introduction of the revised index $a_0$. Because of the
limitation of the normalization, changing the bending degree in
low-$p_T$ region will change the slope in high-$p_T$ region.
Although writing $Cm_T^{a_0}$ in Eq. (5) is not ideal, as it
yields a fractional power unit in $C$, we have no suitable method
to scale out the unit by e.g. $m_0$ due to the nonlinear
relationship between $m_T$ and $m_0$ shown in Eq. (2). In Eq. (5),
the other parameters such as $q$ and $a_0$ do not appear in the
function name for the purpose of convenience. In this work, we
call Eq. (5) the revised Tsallis-like function.

In the framework of the multisource thermal model~\cite{32a}, we
assume that two participant partons take part in the collisions.
Let $p_{t1}$ and $p_{t2}$ denote the components contributed by the
first and second participant parton to $p_T$ respectively, where
$p_{t1}$ ($p_{t2}$) is less than the transverse momentum of the
participant parton. We have
\begin{align}
p_T=\sqrt{p_{t1}^2+p_{t2}^2},
\end{align}
where the two components are perpendicular due to the fact that
$p_{t1}$ and $p_{t2}$ are assumed to be the two components of the
vector $\mathbf{p_T}$. Although multiparton collisions can be
important especially for central high-energy nucleus-nucleus
collisions, the main contributors to particle production are still
binary parton collisions, which are also the basic collision
process. After all, the probability that three or more partons
collide simultaneously is small. Instead, the probability of
binary parton collisions is large.

In binary parton collisions, each parton, e.g. the $i$-th parton,
is assumed to contribute to $p_T$ to obey Eq. (5), where $i=1$ and
2. The probability density functions at mid-$y$ obeyed by $p_{t1}$
and $p_{t2}$ is
\begin{align}
f_i(p_{ti},T) &= C m_{ti}^{a_0}
\bigg[1+\frac{(q-1)(m_{ti}-\mu_i-m_{0i})}{T}\bigg]^{-1/(q-1)},
\end{align}
where the subscript $i$ is used for the quantities related to the
$i$-th parton and $m_{0i}$ is empirically the constituent mass of
the considered parton. Generally, in the case of considering $u$
and/or $d$ quarks, we take $m_u=m_d=0.3$ GeV/$c^2$. It is noted
that the constituent quark masses of 0.3 GeV are not incompatible
with the pion and kaon masses because the collisions between the
two participant quarks can produce more than one particle. The
conservation of energy is satisfied in the collisions. The value
of $\mu_i$ will be discussed at the end of this section.

Let $\phi$ denote the azimuthal angle of $p_T$ relative to
$p_{t1}$. According to refs.~\cite{34,35}, we have the unit
normalized probability density function of $p_T$ and $\phi$ to be
\begin{align}
f_{p_T,\phi}(p_T,\phi,T) &= p_Tf_{1,2}(p_{t1},p_{t2},T) \nonumber\\
&= p_Tf_1(p_{t1},T)f_2(p_{t2},T) \nonumber\\
&= p_Tf_1(p_T\cos\phi,T)f_2(p_T\sin\phi,T),
\end{align}
where $f_{1,2}(p_{t1},p_{t2},T)$ denotes the united probability
density function of $p_{t1}$ and $p_{t2}$. Further, we have the
probability density function of $p_T$ to be
\begin{align}
f_{p_T}(p_T,T)&=\int_{0}^{2\pi}f_{p_T,\phi}(p_T,\phi,T)d\phi \nonumber\\
&=p_T\int_{0}^{2\pi}f_1(p_T\cos\phi,T)f_2(p_T\sin\phi,T)d\phi.
\end{align}

Equation (9) can be used to fit the $p_T$ spectra and obtain the
parameters $T$, $q$, and $a_0$. In the case of fitting a wide
$p_T$ spectrum e.g. $p_T>5$ GeV/$c$, Eq. (9) cannot fit well the
spectra in high-$p_T$ region. Then, we need a superposition of one
Eq. (9) with low $T$ and another Eq. (9) with high $T$ to fit the
whole $p_T$ spectrum. As will be seen in Fig. 3(e) in the next
section, the contribution fraction of the low $T$ component is
very large ($\approx99.9\%$). In most cases in Figs. 1--3, we do
not need the superposition due to narrow $p_T$ spectra. In the
case of using a two-component distribution, we have the
probability density function of $p_T$ to be
\begin{align}
f_{p_T}(p_T)=kf_{p_T}(p_T,T_1)+(1-k)f_{p_T}(p_T,T_2),
\end{align}
where $k$ ($1-k$) denotes the contribution fraction of the first
(second) component and $f_{p_T}(p_T,T_1)$ [$f_{p_T}(p_T,T_2)$] is
given by Eq. (9). The second component is related to the
core-corona picture as mentioned later on in detail in subsection
3.3. Correspondingly, the temperature
\begin{align}
T=kT_1+(1-k)T_2
\end{align}
is averaged by weighting the two fractions. The temperature $T$
defined by Eq. (11) reflects the common effective temperature of
the two components which are assumed to stay in a new equilibrium
in which $T$ still characterizes the average kinetic energy.
Similarly, the weighted average can be performed for other
parameters in the two components in Eq. (10).

It should be noted that the limit of the first and second (low-
and high-$p_T$) components is determined by a convenient
treatment. Generally, the contribution fraction $k$ of the first
component should be taken as largely as possible. As will be seen
in the next section, we take $k=1$ in most cases; only in Fig.
3(e) we take $k=0.999$. Because the contribution fraction of the
second component is zero or small enough, Eq. (10) becomes Eq.
(9), and the weighted average of the two parameters in Eq. (10)
becomes the parameter in Eq. (9). Because Eqs. (1), (4), (5), and
(7) are suitable at mid-$y$, Eqs. (8)--(10) are also suitable at
mid-$y$. In addition, the rapidity ranges quoted in the next
section are narrow and around 0, though the concrete ranges are
different. This means that the mentioned equations are applicable.

We would like to point out that although the model used by itself
is not enough to provide information of the deconfinement phase
transition from hadronic matter to QGP, the excitation function of
extracted parameter is expected to show some particular
tendencies. These particular tendencies include, but are not
limited to, the peak and valley structures, the fast and slow
variations, the positive and negative changes, etc. These
particular tendencies are related to the equation of state (EOS)
of the considered matter. The change of EOS reflects the possible
change of interaction mechanism from hadron-dominated to
parton-dominated intermediate state. Then, the deconfinement phase
transition of the considered matter from hadronic matter to QGP is
possible related to the particular tendencies. It is natural that
the explanations are not only for a given set of data. The present
model will show a method to fit and explain the data.

To obtain $\beta_T$, we need to know the slope of $\langle
p_T\rangle$ versus $\overline{m}$ in the source rest frame of the
considered particle. That is, we need to calculate $\langle
p_T\rangle$ and $\overline{m}$. According to Eq. (10), we have
\begin{align}
\langle p_T\rangle=\int_0^{p_{T\max}} p_T f_{p_T}(p_T)dp_T
\end{align}
due to
\begin{align}
\int_0^{p_{T\max}} f_{p_T}(p_T)dp_T=1,
\end{align}
where $p_{T\max}$ denotes the maximum $p_T$.

As the mean energy, $\overline{E}=\overline{m}= \langle
\sqrt{p^2+m_0^2}\rangle$, where $p$ is the momentum of the
considered particle in the source rest frame. The analytical
calculation of $\overline{m}$ is complex. Instead, we can perform
the calculation by the Monte Carlo method. Let $R_{1,2}$ denote
random numbers distributed evenly in $[0,1]$. Each concrete $p_T$
satisfies
\begin{align}
\int_0^{p_T} f_{p_T}(p'_T,T)dp'_T <R_1< \int_0^{p_T+\delta p_T}
f_{p_T}(p'_T,T)dp'_T,
\end{align}
where $\delta p_T$ denotes a small shift relative to $p_T$. Each
concrete emission angle $\theta$ satisfies
\begin{align}
\theta=2\arcsin\sqrt{R_2}
\end{align}
due to the fact that the particle is assumed to be emitted
isotropically in the source rest frame. Each concrete momentum $p$
and energy $E$ can be obtained by
\begin{align}
p=p_T\csc\theta
\end{align}
and
\begin{align}
E=\sqrt{p^2+m_0^2}
\end{align}
respectively.

After repeating calculations multiple times in the Monte Carlo
method, we can obtain $\overline{E}$, that is $\overline{m}$.
Then, the slope of $\langle p_T\rangle$ versus $\overline{m}$ is
identified as $\beta_T$. Meanwhile, the intercept of $T$ versus
$m_0$ is identified as $T_0$. Here, we emphasize that we have used
the alternative method introduced in section 1 to obtain \(T_0\)
and \(\beta_T\).

It should be noted that in some cases the transverse spectra are
shown in terms of $m_T$, but not $p_T$. To transform the
probability density function $f_{p_T}(p_T,T)$ of $p_T$ to the
probability density function $f_{m_T}(m_T,T)$ of $m_T$, we have
the relation
\begin{align}
f_{p_T}(p_T,T)|dp_T|=f_{m_T}(m_T,T)|dm_T|.
\end{align}
Then, we have
\begin{align}
f_{m_T}(m_T,T)=\frac{m_T}{\sqrt{m_T^2-m_0^2}}f_{p_T}\Big(\sqrt{m_T^2-m_0^2},T\Big)
\end{align}
due to Eq. (2). Using the parameters from $m_T$ spectra, we may
also obtain $T_0$, $\langle p_T\rangle$, $\overline{m}$, and
$\beta_T$.

We now discuss the chemical potential $\mu_i$ of the $i$-th
parton. Generally, the chemical potential $\mu$ of a particle
affects obviously the particle production at low
energy~\cite{48a,48b,48c,48d,48e,48f,48g}. For baryons (mostly
protons and neutrons), the chemical potential $\mu_B$ related to
collision energy $\sqrt{s_{NN}}$ is empirically given by
\begin{align}
\mu_B=\frac{1.303}{1+0.286\sqrt{s_{NN}}},
\end{align}
where both $\mu_B$ and $\sqrt{s_{NN}}$ are in the units of
GeV~\cite{48h,48i,48j}. According to ref.~\cite{48a}, we have
$\mu_u=\mu_d=\mu_B/3$ because a proton or neutron consists of
three $u/d$ quarks (i.e. $uud$ or $udd$).
\\

\section{Results and discussion}

\subsection{Comparison with data and tendencies of free
parameters}

\begin{figure*}[!htb]
\centering
\includegraphics[width=13.0cm]{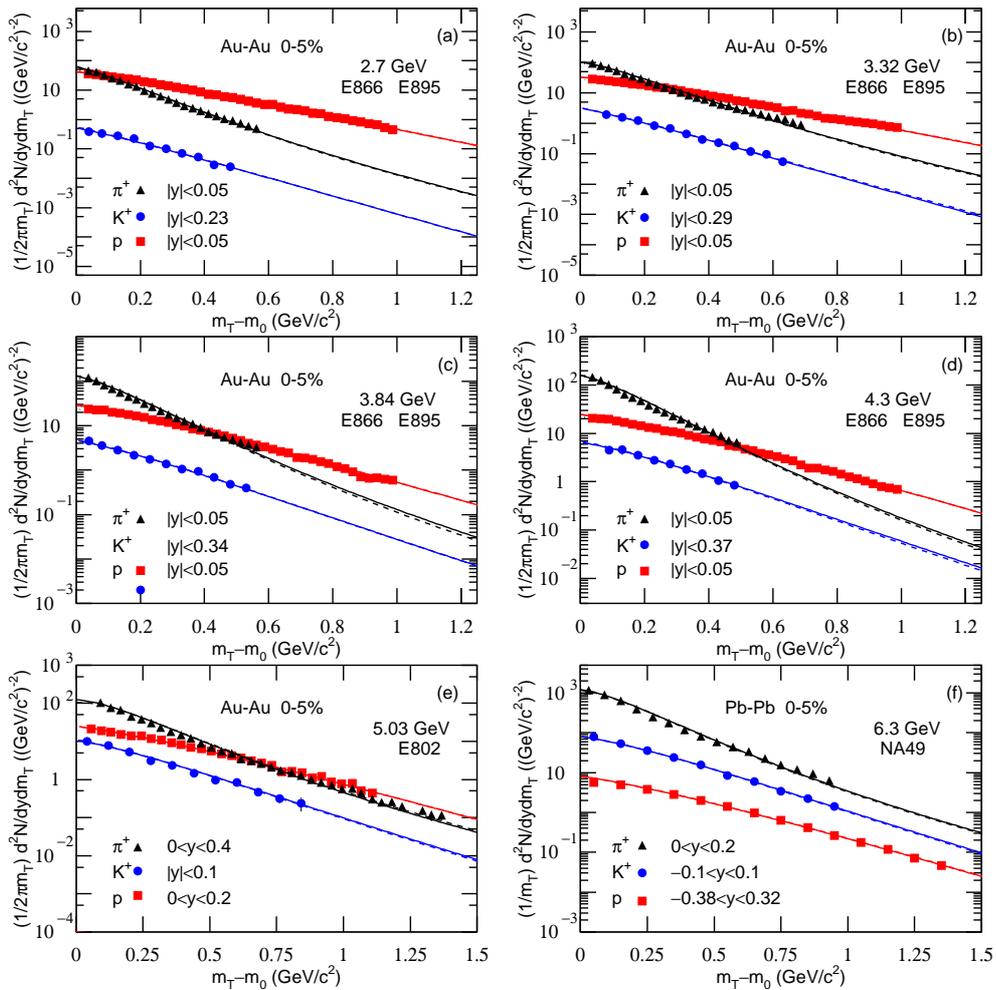}
\caption{Transverse mass spectra of charged pions, kaons, and
protons produced in 0--5\% Au-Au collisions at $\sqrt{s_{NN}}=$
(a) 2.7, (b) 3.32, (c) 3.84, (d) 4.3, and (e) 5.03 GeV, and in
0--5\% Pb-Pb collisions at $\sqrt{s_{NN}}=$ (f) 6.3 GeV. In panel
(f), the factor $1/2\pi$ does not appear, which causes different
normalization from other panels. The symbols represent the
experimental data at mid-$y$ measured by the E866, E895, and E802
Collaboration at the AGS~\cite{E866,E8951,E8952,E802,E8022} and by
the NA49 Collaboration at the SPS~\cite{NA49,NA492}. The solid and
dashed curves are our results, fitted by using Eq. (10) due to
Eqs. (7) and (9), with $\mu_i=0$ and $\mu_i=\mu_B/3$,
respectively.} \label{F1}
\end{figure*}

Figures 1--3 present the transverse momentum $p_T$ (transverse
mass $m_T$) spectra, $(2\pi p_T)^{-1} d^2N/dydp_T$ $[(2\pi
m_T)^{-1} d^2N/dydm_T]$, of charged pions, kaons, and protons
produced in 0--5\% Au-Au, Pb-Pb, and Xe-Xe collisions at different
$\sqrt{s_{NN}}$. The collision types, particle types, mid-$y$
ranges, centrality classes, and $\sqrt{s_{NN}}$ are marked in the
panels. The symbols represent the experimental data measured by
different collaborations. The solid and dashed curves are our
results, fitted by using Eq. (10) due to Eqs. (7) and (9), with
$\mu_i=0$ and $\mu_i=\mu_B/3$, respectively. In the process of
fitting the data, we determine the best parameters by the method
of least squares. The experimental uncertainties used in
calculating the $\chi^2$ are obtained by the root sum square of
the statistical uncertainties and the systematic uncertainties.
The parameters that minimize the $\chi^2$ are the best parameters.
The errors of parameters are obtained by the statistical
simulation method~\cite{48K,48L} which uses the same algorithm as
usual, if not the same Code, in which the errors are also
extracted from variations of $\chi^2$. The values of $T_1$, $T_2$,
$k$, $q$, and $a_0$ are listed in Tables 1 and 2 with the
normalization constant ($N_0$), $\chi^2$, and the number of degree
of freedom (ndof), or explained in the caption of Table 1.

\begin{figure*}[!htb]
\centering
\includegraphics[width=13.0cm]{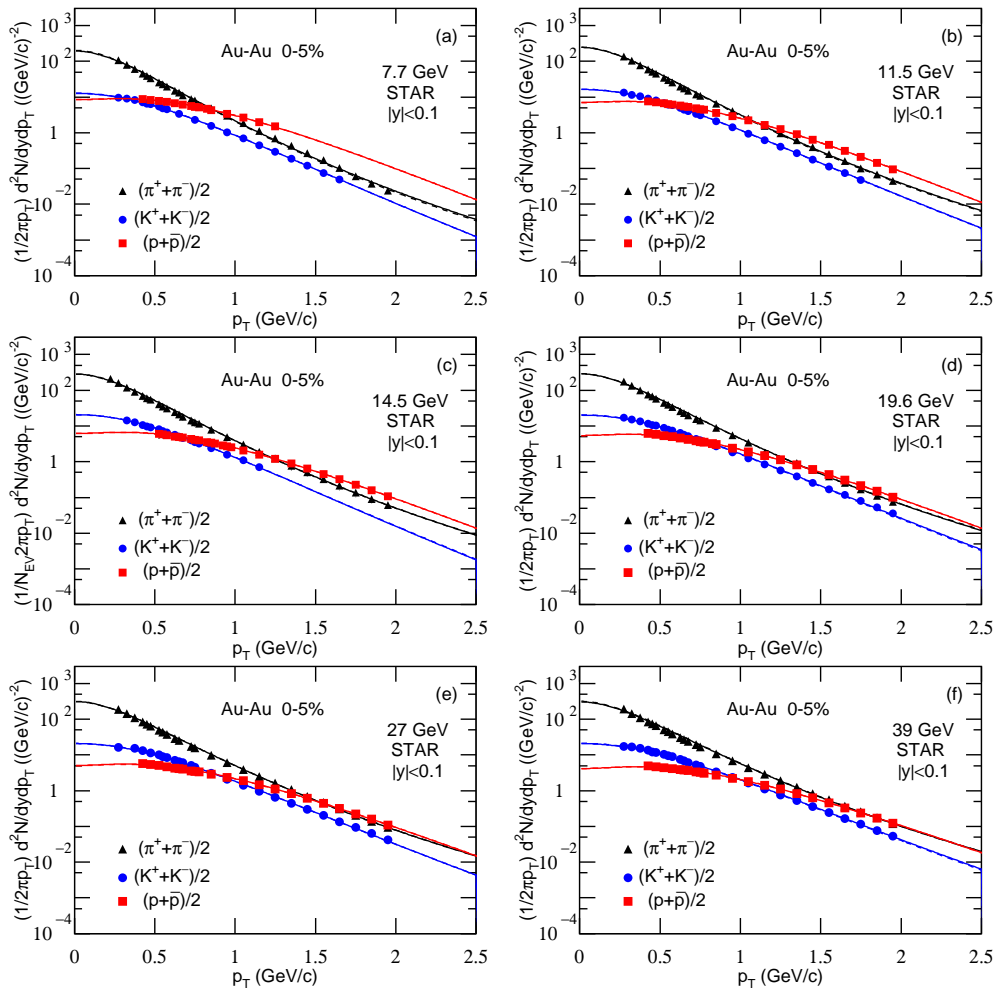}
\caption{Transverse momentum spectra of charged pions, kaons, and
protons produced in 0--5\% Au-Au collisions at $\sqrt{s_{NN}}=$
(a) 7.7, (b) 11.5, (c) 14.5, (d) 19.6, (e) 27, and (f) 39 GeV. In
panel (c), the factor $1/N_{EV}$ i.e. the number of events is
included on the vertical axis, which can be omitted. The symbols
represent the experimental data at mid-$y$ measured by the STAR
Collaboration at the RHIC~\cite{1,2,3}. The solid and dashed
curves are our results, fitted by using Eq. (10) due to Eqs. (7)
and (9), with $\mu_i=0$ and $\mu_i=\mu_B/3$, respectively.}
\label{F2}
\end{figure*}

In a few cases, the values of $\chi^2$/ndof are very large (5--10
or above), which means ``bad" fit to the data. In most cases, the
fits are good due to small $\chi^2$/ndof which is around 1. To
avoid possible wrong interpretation with this result, the number
of ``bad" fits are limited to be much smaller than that of good
fits, for example, 1 to 5 or more strict such as 1 to 10.
Meanwhile, we should also use other method to check the quality of
fits. In fact, we have also calculated the p-values in the Pearson
method. It is shown that all p-values are less than
$3\times10^{-7}$. These p-values corresponds approximately to the
Bayes factor being above 100 and to the confidence degree of
99.99994\% at around 5 standard deviations ($5\sigma$) of the
statistical significance. This means that the model function is in
agreement with the data very well. To say the least, most fits are
acceptable.

It should be noted that we will use a set of pion, kaon, and
proton spectra to extract $T_0$ and $\beta_T$ in subsection 3.2.
For energies in the few GeV range, the spectra of some negative
particles are not available in the literature. So, we have to give
up to analyze all the negative particle spectra in Fig. 1. In our
recent work~\cite{29}, the positive and partial negative particle
spectra were analyzed by the standard distribution. The tendencies
of parameters are approximately independent of isospin, if not the
same for different isospins.

\begin{figure*}[!htb]
\centering
\includegraphics[width=13.0cm]{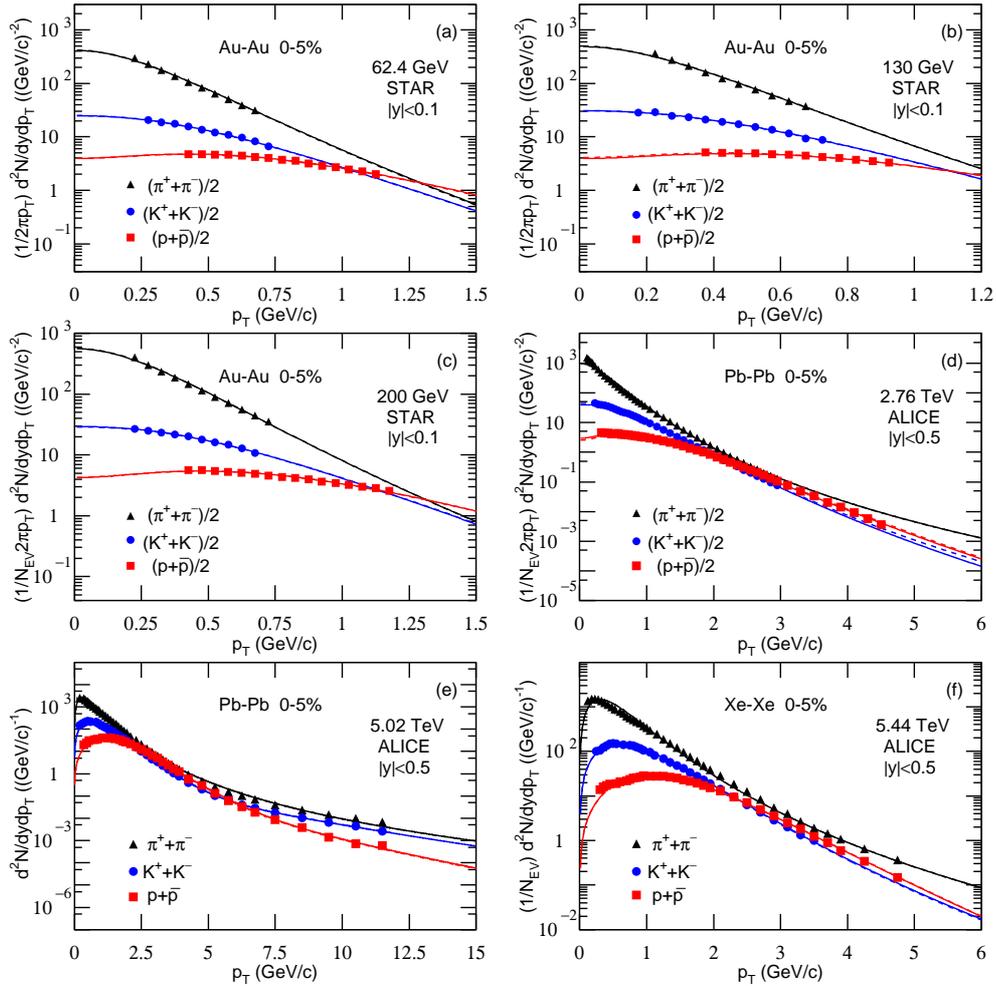}
\caption{Transverse momentum spectra of charged pions, kaons, and
protons produced in 0--5\% Au-Au collisions at $\sqrt{s_{NN}}=$
(a) 62.4, (b) 130, and (c) 200 GeV, in 0--5\% Pb-Pb collisions at
$\sqrt{s_{NN}}=$ (d) 2.76 and (e) 5.02 TeV, and in 0--5\% Xe-Xe
collisions at $\sqrt{s_{NN}}=$ (f) 5.44 TeV. In panels (c), (d),
and (f), the factor $1/N_{EV}$ is included on the vertical axis,
which can be omitted. In panels (e) and (f), the item $(2\pi
p_T)^{-1}$ is not included on the vertical axis, which results in
different calculation for vertical values from other panels in the
normalization. The symbols represent the experimental data at
mid-$y$ measured by the STAR Collaboration at the
RHIC~\cite{1,2,3} and by the ALICE Collaboration at the
LHC~\cite{4,5,6}. The solid and dashed curves are our results,
fitted by using Eq. (10) due to Eqs. (7) and (9), with $\mu_i=0$
and $\mu_i=\mu_B/3$, respectively.} \label{F3}
\end{figure*}

One can see from Figs. 1--3 and Tables l and 2 that Eq. (10)
describes approximately the considered experimental data. For all
energies and particles $T_1$ and $T_2$ are identical except for
the 5.02 TeV Pb-Pb data from ALICE. This means that none of the
spectra have a wide enough range to determine the second component
except the data at 5.02 TeV. The two-component fit is only really
used at 5.02 TeV. In the high-$p_T$ region, the hard scattering
process which is described by the second component in Eq. (10)
contributes totally. However, in the case of using the
two-component function, $k$ ($=0.999$) is very close to 1, which
implies that the contribution of the second component is
negligible. In fact, the second component contributes to the
spectrum in high-$p_T$ region with small fraction, which does not
affect significantly the extraction of parameters. Instead, the
parameters are determined mainly by the spectrum in low-$p_T$
region.

\begin{table*}[htbp]
\caption{Values of free parameters ($T_1$, $T_2$, $q$, and $a_0$),
normalization constant ($N_0$), $\chi^2$, and ndof corresponding
to the solid curves in Figs. 1--3 in which the data are measured
in special conditions (mid-$y$ ranges and energies) by different
collaborations, where $T_2$ is not available in most cases because
$k=1$. In a few cases (at $\sqrt{s_{NN}}=5.02$ TeV), $T_2$ is
available in the next line, where $k=0.999\pm0.001$ which is not
listed in the table.}{\scriptsize
\begin{center}
\begin{tabular} {cccccccc}\\ \hline Collab. & $\sqrt{s_{NN}}$ (GeV) &
Particle & $T_1$, $T_2$ (MeV) & $q$ & $a_0$ & $N_0$ & $\chi^2$/ndof \\
\hline
E866/E895 & 2.7  & $\pi^+$       & $130\pm4$  & $1.062\pm0.003$ & $-0.60\pm0.01$ & $12\pm2$        & 11.87/19\\
Au-Au     &~     & $K^+$         & $143\pm7$  & $1.009\pm0.004$ & $0.49\pm0.01$  & $0.054\pm0.002$ & 3.61/6\\
          &~     & $p$           & $183\pm4$  & $1.005\pm0.001$ & $1.52\pm0.01$  & $75\pm6$        & 153.83/36\\
          & 3.32 & $\pi^+$       & $148\pm4$  & $1.073\pm0.003$ & $-0.53\pm0.01$ & $28\pm2$        & 56.96/24\\
          &~     & $K^+$         & $147\pm6$  & $1.010\pm0.003$ & $0.48\pm0.02$  & $2.14\pm0.01$   & 2.23/8\\
          &~     & $p$           & $194\pm5$  & $1.005\pm0.002$ & $1.67\pm0.01$  & $69\pm3$        & 237.28/36\\
          & 3.84 & $\pi^+$       & $153\pm4$  & $1.075\pm0.003$ & $-0.51\pm0.02$ & $37\pm6$        & 34.34/19\\
          &~     & $K^+$         & $165\pm8$  & $1.022\pm0.005$ & $0.68\pm0.02$  & $4.52\pm0.01$   & 0.92/7\\
          &~     & $p$           & $195\pm5$  & $1.005\pm0.002$ & $1.64\pm0.02$  & $61\pm5$        & 308.11/36\\
          & 4.3  & $\pi^+$       & $155\pm6$  & $1.077\pm0.003$ & $-0.49\pm0.02$ & $46\pm9$        & 47.97/16\\
          &~     & $K^+$         & $172\pm10$ & $1.026\pm0.002$ & $0.72\pm0.02$  & $7.17\pm0.02$   & 0.62/5\\
          &~     & $p$           & $202\pm7$  & $1.007\pm0.003$ & $1.72\pm0.02$  & $59\pm9$        & 75.53/36\\
\hline
E802  & 5.03 & $\pi^+$         & $175\pm4$  & $1.087\pm0.001$ & $-0.37\pm0.01$ & $54\pm6$        & 129.94/30\\
Au-Au &~     & $K^+$             & $174\pm6$  & $1.025\pm0.001$ & $0.71\pm0.01$  & $12\pm3$        & 5.29/7\\
      &~     & $p$               & $206\pm7$  & $1.007\pm0.003$ & $1.79\pm0.03$  & $62\pm5$        & 47.59/25\\
\hline
NA49  & 6.3  & $\pi^+$          & $175\pm4$  & $1.090\pm0.001$ & $-0.45\pm0.02$ & $74\pm6$        & 314.01/12\\
Pb-Pb &~     & $K^+$             & $182\pm6$  & $1.025\pm0.002$ & $0.77\pm0.01$  & $100\pm2$       & 41.92/6\\
      &~     & $p$               & $202\pm7$  & $1.007\pm0.003$ & $1.72\pm0.03$  & $20\pm1$        & 6.99/10\\
\hline
STAR  & 7.7  & $(\pi^++\pi^-)/2$ & $180\pm7$  & $1.077\pm0.001$ & $-0.27\pm0.01$ & $90\pm2$        & 42.38/22\\
Au-Au &~     & $(K^++K^-)/2$     & $189\pm9$  & $1.025\pm0.005$ & $1.00\pm0.01$  & $14\pm3$        & 3.02/16\\
      &~     & $(p+\bar{p})/2$   & $216\pm10$ & $1.007\pm0.002$ & $1.82\pm0.01$  & $27\pm1$        & 0.95/11\\
      & 11.5 & $(\pi^++\pi^-)/2$ & $184\pm7$  & $1.080\pm0.001$ & $-0.23\pm0.01$ & $120\pm5$       & 44.84/22\\
      &~     & $(K^++K^-)/2$     & $192\pm9$  & $1.028\pm0.003$ & $0.98\pm0.01$  & $19\pm3$        & 1.07/19\\
      &~     & $(p+\bar{p})/2$   & $216\pm11$ & $1.007\pm0.001$ & $1.80\pm0.01$  & $23\pm1$        & 1.38/19\\
      & 14.5 & $(\pi^++\pi^-)/2$ & $186\pm7$  & $1.082\pm0.001$ & $-0.23\pm0.02$ & $142\pm9$       & 4.09/24\\
      &~     & $(K^++K^-)/2$     & $189\pm9$  & $1.024\pm0.006$ & $0.97\pm0.01$  & $22\pm3$        & 0.84/14\\
      &~     & $(p+\bar{p})/2$   & $220\pm12$ & $1.010\pm0.001$ & $1.80\pm0.01$  & $21\pm1$        & 0.28/21\\
      & 19.6 & $(\pi^++\pi^-)/2$ & $189\pm8$  & $1.086\pm0.001$ & $-0.25\pm0.03$ & $150\pm6$       & 32.66/21\\
      &~     & $(K^++K^-)/2$     & $200\pm9$  & $1.026\pm0.003$ & $0.98\pm0.01$  & $24\pm4$        & 19.01/22\\
      &~     & $(p+\bar{p})/2$   & $222\pm11$ & $1.011\pm0.001$ & $1.80\pm0.02$  & $19\pm1$        & 2.20/18\\
      & 27   & $(\pi^++\pi^-)/2$ & $191\pm8$  & $1.089\pm0.001$ & $-0.22\pm0.01$ & $164\pm6$       & 27.71/21\\
      &~     & $(K^++K^-)/2$     & $202\pm9$  & $1.027\pm0.003$ & $0.99\pm0.01$  & $26\pm3$        & 10.49/20\\
      &~     & $(p+\bar{p})/2$   & $225\pm11$ & $1.011\pm0.002$ & $1.81\pm0.02$  & $19\pm1$        & 4.56/18\\
      & 39   & $(\pi^++\pi^-)/2$ & $196\pm9$  & $1.091\pm0.001$ & $-0.16\pm0.03$ & $170\pm9$       & 35.77/22\\
      &~     & $(K^++K^-)/2$     & $207\pm10$ & $1.031\pm0.002$ & $0.97\pm0.01$  & $28\pm3$        & 9.02/22\\
      &~     & $(p+\bar{p})/2$   & $232\pm12$ & $1.012\pm0.001$ & $1.82\pm0.01$  & $17\pm2$        & 1.64/18\\
      & 62.4 & $(\pi^++\pi^-)/2$ & $189\pm9$  & $1.078\pm0.001$ & $-0.25\pm0.02$ & $208\pm9$       & 103.95/6\\
      &~     & $(K^++K^-)/2$     & $212\pm10$ & $1.031\pm0.001$ & $0.99\pm0.01$  & $35\pm3$        & 1.50/6\\
      &~     & $(p+\bar{p})/2$   & $243\pm13$ & $1.020\pm0.002$ & $1.88\pm0.02$  & $22\pm1$        & 5.98/11\\
      & 130  & $(\pi^++\pi^-)/2$ & $190\pm9$  & $1.078\pm0.002$ & $-0.26\pm0.01$ & $245\pm9$       & 122.72/6\\
      &~     & $(K^++K^-)/2$     & $213\pm10$ & $1.031\pm0.003$ & $1.00\pm0.01$  & $44\pm3$        & 2.23/8\\
      &~     & $(p+\bar{p})/2$   & $247\pm13$ & $1.021\pm0.002$ & $1.87\pm0.02$  & $23\pm1$        & 20.75/8\\
      & 200  & $(\pi^++\pi^-)/2$ & $192\pm9$  & $1.080\pm0.003$ & $-0.26\pm0.01$ & $286\pm9$       & 85.21/7\\
      &~     & $(K^++K^-)/2$     & $218\pm11$ & $1.034\pm0.002$ & $1.11\pm0.02$  & $49\pm3$        & 0.42/6\\
      &~     & $(p+\bar{p})/2$   & $250\pm14$ & $1.024\pm0.002$ & $1.93\pm0.01$  & $28\pm1$        & 27.56/12\\
\hline
ALICE & 2760 & $(\pi^++\pi^-)/2$ & $230\pm10$ & $1.140\pm0.001$ & $-0.16\pm0.00$ & $709\pm11$      & 155.11/37\\
Pb-Pb &~     & $(K^++K^-)/2$     & $251\pm13$ & $1.067\pm0.002$ & $1.09\pm0.02$  & $109\pm6$       & 4.63/32\\
      &~     & $(p+\bar{p})/2$   & $300\pm14$ & $1.043\pm0.001$ & $1.86\pm0.03$  & $32\pm3$        & 22.39/38\\
      & 5020 & $\pi^++\pi^-$     & $231\pm11$ & $1.138\pm0.001$ & $-0.15\pm0.01$ & $1899\pm30$     & 153.36/36\\
      &      &                   & $999\pm18$ &                 &                &                 &        \\
      &~     & $ K^++K^-$        & $250\pm13$ & $1.067\pm0.001$ & $1.21\pm0.01$  & $269\pm10$      & 5.95/32\\
      &      &                   & $1100\pm20$&                 &                &                 &        \\
      &~     & $p+\bar{p}$       & $321\pm14$ & $1.045\pm0.001$ & $1.77\pm0.02$  & $72\pm4$        & 19.51/27\\
      &      &                   & $999\pm16$ &                 &                &                 &        \\
\hline
ALICE & 5440 & $\pi^++\pi^-$     & $238\pm12$ & $1.140\pm0.002$ & $-0.15\pm0.01$ & $1057\pm33$     & 21.89/36\\
Xe-Xe &~     & $K^++K^-$         & $260\pm13$ & $1.068\pm0.002$ & $1.08\pm0.02$  & $168\pm11$      & 1.49/27\\
      &~     & $p+\bar{p}$       & $327\pm14$ & $1.040\pm0.001$ & $1.71\pm0.04$  & $49\pm3$        & 11.75/30\\
\hline
\end{tabular}%
\end{center}}
\end{table*}

\begin{table*}[!htb]
\caption{Values of $T_1$, $T_2$, $q$, $a_0$, $N_0$, $\chi^2$, and
ndof corresponding to the dashed curves in Figs. 1--3.}
{\scriptsize
\begin{center}
\begin{tabular} {cccccccc}\\ \hline Collab. & $\sqrt{s_{NN}}$ (GeV) &
Particle & $T_1$, $T_2$ (MeV) & $q$ & $a_0$ & $N_0$ & $\chi^2$/ndof \\
\hline
E866/E895 & 2.7  & $\pi^+$       & $139\pm4$  & $1.069\pm0.003$ & $-0.49\pm0.01$ & $12\pm2$        & 12.31/19 \\
Au-Au     &~     & $K^+$         & $145\pm7$  & $1.010\pm0.004$ & $0.46\pm0.01$  & $0.056\pm0.001$ & 3.77/6 \\
          &~     & $p$           & $183\pm4$  & $1.005\pm0.001$ & $1.55\pm0.01$  & $76\pm6$        & 148.48/36 \\
          & 3.32 & $\pi^+$       & $159\pm4$  & $1.078\pm0.003$ & $-0.45\pm0.01$ & $28\pm2$        & 62.79/24 \\
          &~     & $K^+$         & $150\pm6$  & $1.013\pm0.003$ & $0.47\pm0.02$  & $2.19\pm0.01$   & 2.14/8 \\
          &~     & $p$           & $194\pm5$  & $1.005\pm0.002$ & $1.66\pm0.01$  & $69\pm3$        & 244.84/36 \\
          & 3.84 & $\pi^+$       & $159\pm4$  & $1.077\pm0.003$ & $-0.42\pm0.02$ & $37\pm6$        & 45.43/19 \\
          &~     & $K^+$         & $168\pm8$  & $1.023\pm0.005$ & $0.69\pm0.02$  & $4.59\pm0.01$   & 0.94/7 \\
          &~     & $p$           & $195\pm5$  & $1.005\pm0.002$ & $1.64\pm0.02$  & $61\pm5$        & 310.66/36 \\
          & 4.3  & $\pi^+$       & $162\pm6$  & $1.080\pm0.003$ & $-0.42\pm0.02$ & $46\pm9$        & 56.47/16 \\
          &~     & $K^+$         & $173\pm10$ & $1.026\pm0.002$ & $0.72\pm0.02$  & $7.20\pm0.02$   & 0.81/5 \\
          &~     & $p$           & $202\pm7$  & $1.007\pm0.003$ & $1.74\pm0.02$  & $59\pm9$        & 74.66/36 \\
\hline
E802  & 5.03 & $\pi^+$           & $183\pm4$  & $1.092\pm0.001$ & $-0.23\pm0.01$ & $53\pm6$        & 164.90/30 \\
Au-Au &~     & $K^+$             & $175\pm6$  & $1.026\pm0.001$ & $0.72\pm0.01$  & $12\pm3$        & 4.58/7 \\
      &~     & $p$               & $205\pm7$  & $1.007\pm0.003$ & $1.74\pm0.03$  & $62\pm5$        & 65.21/25 \\
\hline
NA49  & 6.3  & $\pi^+$           & $185\pm4$  & $1.093\pm0.001$ & $-0.42\pm0.02$ & $72\pm6$        & 328.26/12 \\
Pb-Pb &~     & $K^+$             & $175\pm6$  & $1.026\pm0.002$ & $0.78\pm0.01$  & $100\pm2$       & 30.95/6 \\
      &~     & $p$               & $205\pm7$  & $1.007\pm0.003$ & $1.73\pm0.03$  & $20\pm1$        & 6.79/10 \\
\hline
STAR  & 7.7  & $(\pi^++\pi^-)/2$ & $185\pm7$  & $1.079\pm0.001$ & $-0.25\pm0.01$ & $91\pm2$        & 54.50/22 \\
Au-Au &~     & $(K^++K^-)/2$     & $190\pm9$  & $1.026\pm0.005$ & $1.03\pm0.01$  & $14\pm3$        & 1.90/16 \\
      &~     & $(p+\bar{p})/2$   & $216\pm10$ & $1.007\pm0.002$ & $1.82\pm0.01$  & $27\pm1$        & 1.33/11 \\
      & 11.5 & $(\pi^++\pi^-)/2$ & $187\pm7$  & $1.083\pm0.001$ & $-0.21\pm0.01$ & $120\pm5$       & 41.38/22 \\
      &~     & $(K^++K^-)/2$     & $194\pm9$  & $1.029\pm0.003$ & $0.99\pm0.01$  & $19\pm3$        & 1.03/19 \\
      &~     & $(p+\bar{p})/2$   & $216\pm11$ & $1.007\pm0.001$ & $1.82\pm0.01$  & $23\pm1$        & 1.36/19 \\
      & 14.5 & $(\pi^++\pi^-)/2$ & $190\pm7$  & $1.084\pm0.001$ & $-0.20\pm0.02$ & $141\pm9$       & 3.74/24 \\
      &~     & $(K^++K^-)/2$     & $191\pm9$  & $1.025\pm0.006$ & $0.97\pm0.01$  & $22\pm3$        & 0.81/14 \\
      &~     & $(p+\bar{p})/2$   & $220\pm12$ & $1.010\pm0.001$ & $1.82\pm0.01$  & $21\pm1$        & 0.30/21 \\
      & 19.6 & $(\pi^++\pi^-)/2$ & $192\pm8$  & $1.089\pm0.001$ & $-0.18\pm0.03$ & $150\pm6$       & 39.67/21 \\
      &~     & $(K^++K^-)/2$     & $201\pm9$  & $1.026\pm0.003$ & $0.96\pm0.01$  & $24\pm4$        & 17.06/22 \\
      &~     & $(p+\bar{p})/2$   & $222\pm11$ & $1.011\pm0.001$ & $1.81\pm0.02$  & $19\pm1$        & 2.27/18 \\
      & 27   & $(\pi^++\pi^-)/2$ & $193\pm8$  & $1.091\pm0.001$ & $-0.18\pm0.01$ & $164\pm6$       & 27.36/21 \\
      &~     & $(K^++K^-)/2$     & $203\pm9$  & $1.028\pm0.003$ & $0.99\pm0.01$  & $26\pm3$        & 10.01/20 \\
      &~     & $(p+\bar{p})/2$   & $225\pm11$ & $1.011\pm0.002$ & $1.82\pm0.02$  & $19\pm1$        & 4.67/18 \\
      & 39   & $(\pi^++\pi^-)/2$ & $198\pm9$  & $1.091\pm0.001$ & $-0.16\pm0.03$ & $176\pm9$       & 59.05/22 \\
      &~     & $(K^++K^-)/2$     & $208\pm10$ & $1.031\pm0.002$ & $0.97\pm0.01$  & $28\pm3$        & 9.05/22 \\
      &~     & $(p+\bar{p})/2$   & $232\pm12$ & $1.012\pm0.001$ & $1.82\pm0.01$  & $17\pm2$        & 1.65/18 \\
      & 62.4 & $(\pi^++\pi^-)/2$ & $189\pm9$  & $1.078\pm0.001$ & $-0.25\pm0.02$ & $208\pm9$       & 97.82/6 \\
      &~     & $(K^++K^-)/2$     & $212\pm10$ & $1.031\pm0.001$ & $1.00\pm0.01$  & $35\pm3$        & 1.50/6 \\
      &~     & $(p+\bar{p})/2$   & $243\pm13$ & $1.020\pm0.002$ & $1.88\pm0.02$  & $21\pm1$        & 16.62/11 \\
      & 130  & $(\pi^++\pi^-)/2$ & $190\pm9$  & $1.078\pm0.002$ & $-0.26\pm0.01$ & $248\pm9$       & 143.34/6 \\
      &~     & $(K^++K^-)/2$     & $213\pm10$ & $1.031\pm0.003$ & $1.00\pm0.01$  & $44\pm3$        & 2.25/8 \\
      &~     & $(p+\bar{p})/2$   & $247\pm13$ & $1.021\pm0.002$ & $1.87\pm0.02$  & $23\pm1$        & 19.97/8 \\
      & 200  & $(\pi^++\pi^-)/2$ & $192\pm9$  & $1.080\pm0.003$ & $-0.26\pm0.01$ & $288\pm9$       & 111.25/7 \\
      &~     & $(K^++K^-)/2$     & $218\pm11$ & $1.034\pm0.002$ & $1.12\pm0.02$  & $48\pm3$        & 0.42/6 \\
      &~     & $(p+\bar{p})/2$   & $250\pm14$ & $1.024\pm0.002$ & $1.93\pm0.01$  & $28\pm1$        & 28.32/12 \\
\hline
ALICE & 2760 & $(\pi^++\pi^-)/2$ & $230\pm10$ & $1.140\pm0.001$ & $-0.16\pm0.01$ & $709\pm11$      & 155.11/37 \\
Pb-Pb &~     & $(K^++K^-)/2$     & $251\pm13$ & $1.067\pm0.002$ & $1.09\pm0.02$  & $109\pm6$       & 4.64/32 \\
      &~     & $(p+\bar{p})/2$   & $300\pm14$ & $1.043\pm0.001$ & $1.86\pm0.03$  & $32\pm3$        & 22.50/38 \\
      & 5020 & $\pi^++\pi^-$     & $231\pm11$ & $1.138\pm0.001$ & $-0.15\pm0.01$ & $1899\pm30$     & 153.36/36 \\
      &      &                   & $999\pm18$ &                 &                &                 &         \\
      &~     & $ K^++K^-$        & $250\pm13$ & $1.067\pm0.001$ & $1.21\pm0.01$  & $269\pm10$      & 5.94/32 \\
      &      &                   & $1100\pm20$&                 &                &                 &         \\
      &~     & $p+\bar{p}$       & $321\pm14$ & $1.045\pm0.001$ & $1.77\pm0.02$  & $72\pm4$        & 19.49/27 \\
      &      &                   & $999\pm16$ &                 &                &                 &         \\
\hline
ALICE & 5440 & $\pi^++\pi^-$     & $238\pm12$ & $1.140\pm0.002$ & $-0.15\pm0.01$ & $1057\pm33$     & 21.89/36 \\
Xe-Xe &~     & $K^++K^-$         & $260\pm13$ & $1.068\pm0.002$ & $1.08\pm0.02$  & $168\pm11$      & 1.49/27 \\
      &~     & $p+\bar{p}$       & $327\pm14$ & $1.040\pm0.001$ & $1.71\pm0.04$  & $49\pm3$        & 11.74/30 \\
\hline
\end{tabular}%
\end{center}}
\end{table*}

\begin{figure*}[!htb]
\centering
\includegraphics[width=13.0cm]{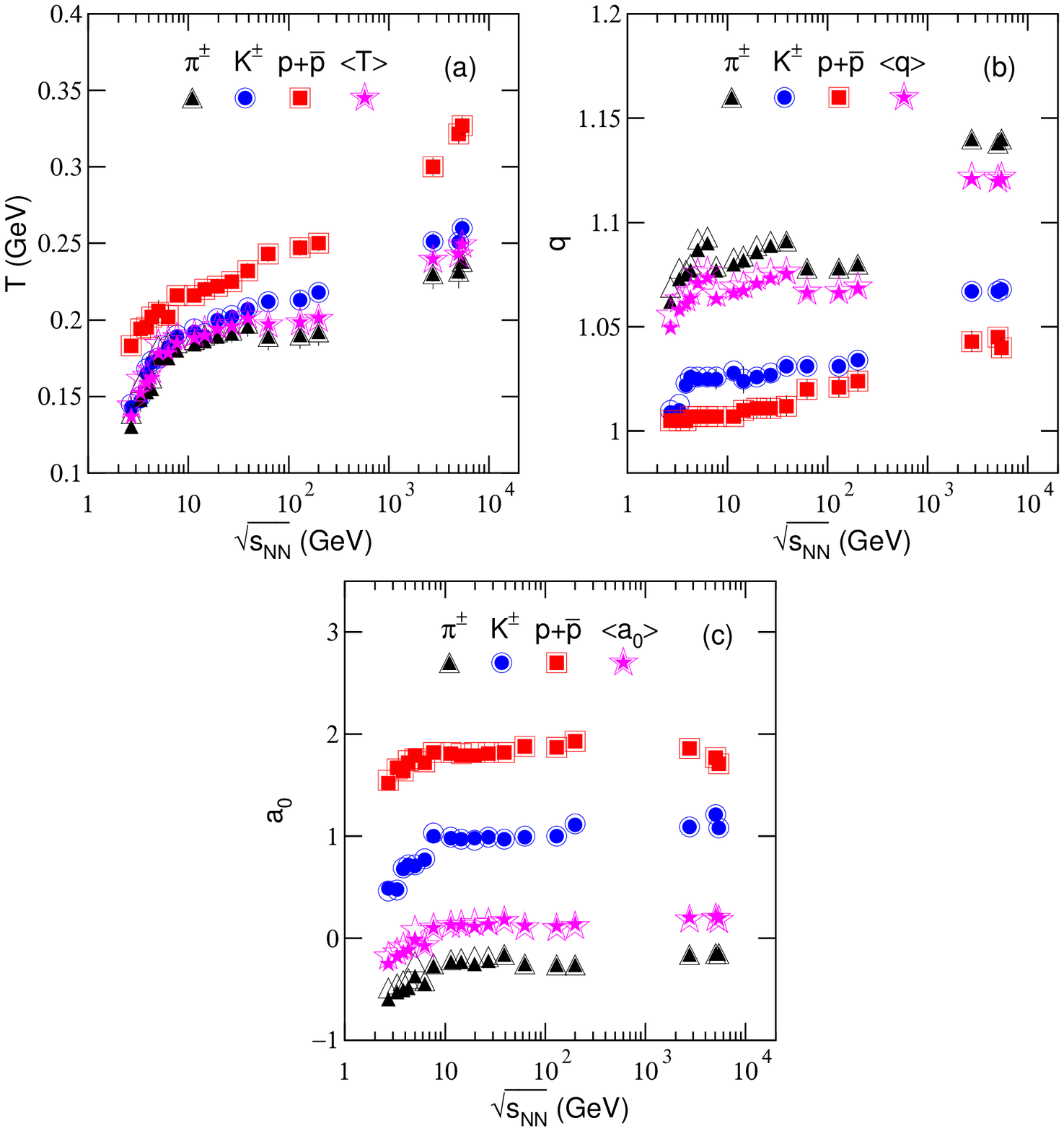}
\caption{Dependences of (a) effective temperature $T$, (b) entropy
index $q$, and (c) revised index $a_0$ on energy $\sqrt{s_{NN}}$,
where the closed and open symbols are cited from Tables 1 and 2
which are obtained from the fittings with $\mu_i=0$ (solid curves)
and $\mu_i=\mu_B/3$ (dashed curves) in Figs 1--3, respectively.
The triangles, circles, squares, and pentagrams represent the
results for charged pions, kaons, protons, and the average by
weighting different yields, respectively.} \label{F4}
\end{figure*}

Although the contribution fraction of the second component is very
small, the spectra with wide $p_T$ range on Fig. 3(e) is well fit
using the two components, it means increasing the number of
parameters compared with just Tsallis function. Generally, the
spectrum shapes of different particles are different. However, we
may use the same function with different parameters and
normalization constants to fit them uniformly. In some cases, the
spectrum forms are different. We need to consider corresponding
normalization treatments so that the fitting function and the data
are compatible and concordant.

The value of $\mu_i$ affects mainly the parameters at below dozens
of GeV. Although $\mu_i=0$ is not justified at lower energies, we
present the results with $\mu_i=0$ for comparison with
$\mu_i=\mu_B/3$ so that we can have a quantitative understanding
on the influence of $\mu_i$. It should be noted that $\mu_i$ is
only for $\mu_u$ and $\mu_d$, that is $\mu_u=\mu_d=\mu_B/3$. For
pions, we have $\mu_{\pi}=\mu_u+\mu_d=2\mu_B/3$. For kaons, we
have no suitable expression because the chemical potential $\mu_s$
for $s$ quark is not available here. Generally, $\mu_s>\mu_u$. So,
$\mu_K=\mu_u+\mu_s>\mu_{\pi}$.

\begin{figure*}[!htb]
\centering
\includegraphics[width=13.0cm]{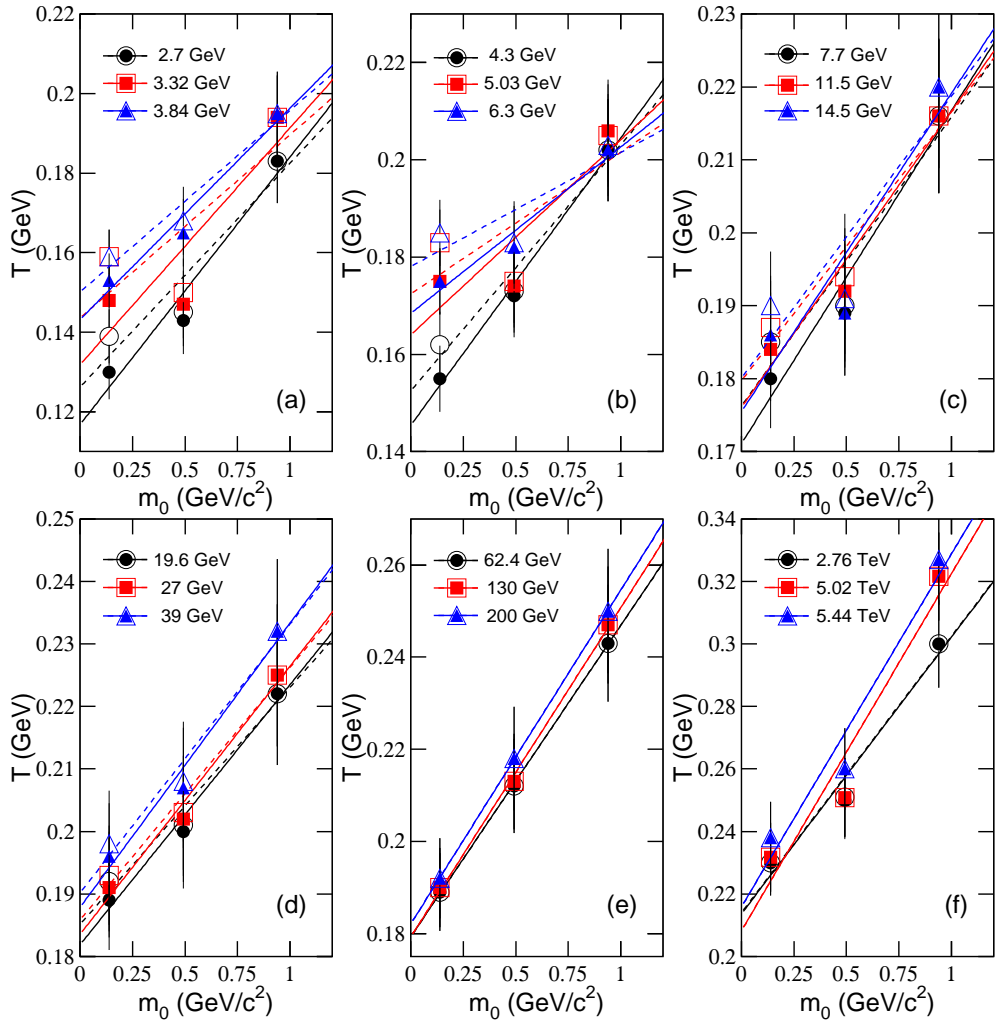}
\caption{Dependences of $T$ on $m_0$. Different symbols represent
the results from identified particles produced in central AA
collisions at different energies shown in panels (a)--(f). The
lines are the results fitted by the least square method, where the
intercepts are regarded as $T_0$. The closed and open symbols (the
solid and dashed curves) correspond to the results for $\mu_i=0$
and $\mu_i=\mu_B/3$ respectively.} \label{F5}
\end{figure*}

As a function with wide applications, the Tsallis distribution can
describe in fact the spectra presented in Figs. 1--3 in most
cases, though the values of parameters may be changed. However, to
extract some information at the parton level, we have regarded the
revised Tsallis-like function [Eq. (7)] as the components of $p_T$
contributed by the participant partons. The value of $p_T$ is then
taken to be the root sum square of the components. In the present
work, we have considered two participant partons and two
components. This treatment can be extended to three and more
participant partons and their components. In the case of the
analytical expression for more components becoming difficult, we
may use the Monte Carlo method to obtain the components, and $p_T$
is also the root sum square of the components. Then, the
distribution of $p_T$ is obtained by the statistical method.

To study the changing trends of the free parameters, Fig. 4 shows
the dependences of (a) effective temperature $T$, (b) entropy
index $q$, and (c) revised index $a_0$ on collision energy
$\sqrt{s_{NN}}$, where the closed and open symbols are cited from
Tables 1 and 2 which are obtained from the fittings with $\mu_i=0$
(solid curves) and $\mu_i=\mu_B/3$ (dashed curves) in Figs. 1--3,
respectively. The triangles, circles, squares, and pentagrams
represent the results for charged pions, kaons, protons, and the
average by weighting different yields, respectively. Because the
errors of parameters are very small, the error bars in the plots
are invisible. One can see from Fig. 4 that, $T$ increases
significantly, $q$ increases slowly, and $a_0$ increase quickly
from $\approx3$ to $\approx10$ GeV (exactly from 2.7 to 7.7 GeV)
and then changes slowly at above 10 GeV except for a large
increase ($\approx50\%$) at the maximum energy, with the increase
of $\ln(\sqrt{s_{NN}})$. These parameters also show their
dependences on particle mass $m_0$: With the increase of $m_0$,
$T$ and $a_0$ increase and $q$ decreases significantly. Indeed,
$\mu_i$ affects only the parameters at the lower energies (below
dozens of GeV), but not higher energy.

\begin{figure*}[!htb]
\centering
\includegraphics[width=13.0cm]{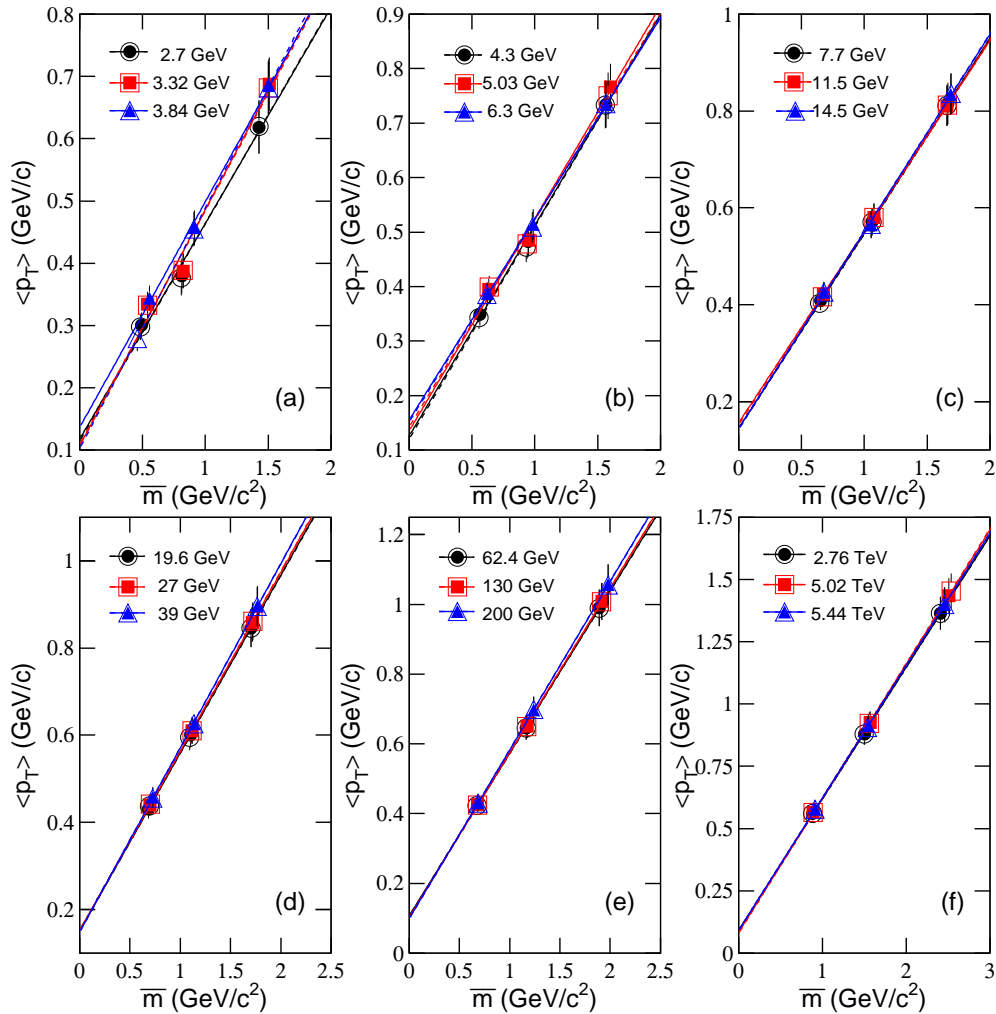}
\caption{Same as for Fig. 5, but showing the dependences of
$\langle p_T\rangle$ on $\overline{m}$. The lines are the results
fitted by the least square method, where the slopes are regarded
as $\beta_T$.} \label{F6}
\end{figure*}

The behaviour of excitation function of $T$ will be discussed as
that of $T_0$ in the next subsection. The large fluctuations of
$q$ for pions are caused by the large influence of strong decay
from high-mass resonance and weak decay from heavy flavor hadrons.
For light particles such as pions, the influence and then the
fluctuations are large; while for relative heavy particles such as
kaons and protons, the influence and then the fluctuations are
small. No matter how large the fluctuations are, the values of $q$
are close to 1.

As we mentioned in the above section, the entropy index $q$
reflects the degree of equilibrium or non-equilibrium of collision
system. Usually, $q=1$ corresponds to an ideal equilibrium state
and $q\gg1$ means a non-equilibrium state. The present work shows
that $q$ is very close to 1 which means that the system stays in
the equilibrium state. Generally, the equilibrium is relative. For
the case of non-equilibrium, we may use the concept of local
equilibrium. If $q$ is not too large, for example, $q\le1.25$ or
$n\ge4$, the collision system is still in equilibrium or local
equilibrium~\cite{36a,60a}. In particular, the system is closer to
the equilibrium when it emits protons at lower energy, comparing
with pions and kaons at higher energy. The reason is that most
protons came from the participant nuclei directly. They have
enough time to reach to the equilibrium in the evolution. At lower
energy, the system is closer to the equilibrium because the
evolution is slower and the system has more time to result in the
equilibrium. From the initial collisions to kinetic freeze-out,
the evolution time is very short. The lower the collision energy
is, the longer the evolution time is.

\begin{table*}[!htb]
\caption{Values of intercepts, slopes, and $\chi^2$ for the solid
lines in Figs. 5 and 6, where ndof = 1 which is not shown in the
table. The units of the intercepts in Figs. 5 and 6 are GeV and
GeV/$c$ respectively. The units of the slopes in Figs. 5 and 6 are
$c^2$ and $c$ respectively.}
\begin{center}
\begin{tabular} {ccccccc}\\ \hline Figure &  Relation  &  System &  $\sqrt{s_{NN}}$ (GeV) &
 Intercept & Slope & $\chi^2$  \\
 \hline
Fig. 5(a) & $T-m_0$ & Au-Au  & 2.7   & $0.117\pm0.002$  & $0.067\pm0.002$  & 1.08 \\
          &~        &~       & 3.32  & $0.132\pm0.001$  & $0.060\pm0.003$  & 4.50 \\
          &~        &~       & 3.84  & $0.143\pm0.002$  & $0.053\pm0.003$  & 0.43 \\
\hline
Fig. 5(b) & $T-m_0$ & Au-Au  & 4.3   & $0.145\pm0.002$  & $0.059\pm0.004$  & 0.14 \\
          &~        &~       & 5.03  & $0.164\pm0.002$  & $0.040\pm0.003$  & 2.14 \\
          &~        & Pb-Pb  & 6.3   & $0.168\pm0.001$  & $0.034\pm0.004$  & 0.24 \\
\hline
Fig. 5(c) & $T-m_0$ & Au-Au  & 7.7   & $0.171\pm0.002$  & $0.046\pm0.003$  & 0.48 \\
          &~        &~       & 11.5  & $0.176\pm0.002$  & $0.041\pm0.003$  & 0.36 \\
          &~        &~       & 14.5  & $0.176\pm0.001$  & $0.044\pm0.004$  & 1.32 \\
\hline
Fig. 5(d) &$T-m_0$  & Au-Au  & 19.6  & $0.182\pm0.003$  & $0.042\pm0.004$  & 0.11 \\
          &~        &~       & 27    & $0.184\pm0.003$  & $0.043\pm0.004$  & 0.13 \\
          &~        &~       & 39    & $0.188\pm0.003$  & $0.046\pm0.004$  & 0.18 \\
\hline
Fig. 5(e) &$T-m_0$  & Au-Au  & 62.4  & $0.179\pm0.003$  & $0.068\pm0.001$  & 0.01 \\
          &~        &~       & 130   & $0.179\pm0.003$  & $0.072\pm0.004$  & 0.03 \\
          &         & Au-Au  & 200   & $0.182\pm0.004$  & $0.073\pm0.004$  & 0.01 \\
\hline
Fig. 5(f) &$T-m_0$  & Pb-Pb  & 2760  & $0.214\pm0.003$  & $0.089\pm0.004$  & 0.45 \\
          &~        &~       & 5020  & $0.208\pm0.003$  & $0.114\pm0.003$  & 1.84 \\
          &~        & Xe-Xe  & 5440  & $0.216\pm0.003$  & $0.113\pm0.003$  & 1.23 \\
\hline
Fig. 6(a) & $\langle p_T\rangle-\overline m$ & Au-Au  & 2.7   & $0.117\pm0.004$  & $0.347\pm0.004$  & 0.93 \\
          &~                                 &        & 3.32  & $0.106\pm0.004$  & $0.379\pm0.005$  & 2.52 \\
          &~                                 &~       & 3.84  & $0.136\pm0.005$  & $0.363\pm0.005$  & 0.22 \\
\hline
Fig. 6(b) & $\langle p_T\rangle-\overline m$ & Au-Au  & 4.3   & $0.125\pm0.004$  & $0.387\pm0.005$  & 0.17 \\
          &~                                 &~       & 5.03  & $0.135\pm0.004$  & $0.390\pm0.005$  & 0.94 \\
          &~                                 & Pb-Pb  & 6.3   & $0.155\pm0.005$  & $0.369\pm0.004$  & 0.06 \\
\hline
Fig. 6(c) & $\langle p_T\rangle-\overline m$ & Au-Au  & 7.7   & $0.145\pm0.005$  & $0.403\pm0.005$  & 0.01 \\
          &~                                 &~       & 11.5  & $0.156\pm0.005$  & $0.395\pm0.007$  & 0.01 \\
          &~                                 &~       & 14.5  & $0.144\pm0.005$  & $0.407\pm0.006$  & 0.16 \\
\hline
Fig. 6(d) & $\langle p_T\rangle-\overline m$ & Au-Au  & 19.6  & $0.150\pm0.004$  & $0.408\pm0.005$  & 0.01 \\
          &~                                 &~       & 27    & $0.152\pm0.004$  & $0.411\pm0.006$  & 0.01 \\
          &~                                 &~       & 39    & $0.148\pm0.004$  & $0.423\pm0.006$  & 0.21 \\
\hline
Fig. 6(e) &$\langle p_T\rangle-\overline m$  & Au-Au  & 62.4  & $0.106\pm0.003$  & $0.467\pm0.006$  & 0.04 \\
          &~                                 &~       & 130   & $0.102\pm0.003$  & $0.472\pm0.008$  & 0.04 \\
          &~                                 &~       & 200   & $0.098\pm0.003$  & $0.484\pm0.008$  & 0.01 \\
\hline
Fig. 6(f) &$\langle p_T\rangle-\overline m$  & Pb-Pb  & 2760  & $0.089\pm0.002$  & $0.528\pm0.006$  & 0.01 \\
          &~                                 &~       & 5020  & $0.082\pm0.002$  & $0.539\pm0.008$  & 0.01 \\
          &~                                 & Xe-Xe  & 5440  & $0.091\pm0.002$  & $0.532\pm0.009$  & 0.01 \\
\hline
\end{tabular}%
\end{center}
\end{table*}

\begin{table*}[!htb]
\caption{Values of intercepts, slopes, and $\chi^2$ for the dashed
lines in Figs. 5 and 6.}
\begin{center}
\begin{tabular} {ccccccc}\\ \hline Figure &  Relation  &  Type &  $\sqrt{s_{NN}}$ (GeV) &
 Intercept & Slope & $\chi^2$  \\
 \hline
Fig. 5(a) & $T-m_0$ & Au-Au  & 2.7   & $0.126\pm0.002$  & $0.056\pm0.002$  & 1.79 \\
          &~        &~       & 3.32  & $0.144\pm0.001$  & $0.046\pm0.003$  & 5.91 \\
          &~        &~       & 3.84  & $0.150\pm0.002$  & $0.046\pm0.003$  & 0.48 \\
\hline
Fig. 5(b) & $T-m_0$ & Au-Au  & 4.3   & $0.152\pm0.002$  & $0.051\pm0.004$  & 0.45 \\
          &~        &~       & 5.03  & $0.172\pm0.002$  & $0.029\pm0.003$  & 3.10 \\
          &~        & Pb-Pb  & 6.3   & $0.178\pm0.001$  & $0.024\pm0.004$  & 0.98 \\
\hline
Fig. 5(c) & $T-m_0$ & Au-Au  & 7.7   & $0.176\pm0.002$  & $0.040\pm0.003$  & 0.75 \\
          &~        &~       & 11.5  & $0.180\pm0.002$  & $0.037\pm0.003$  & 0.32 \\
          &~        &~       & 14.5  & $0.180\pm0.001$  & $0.039\pm0.004$  & 1.37 \\
\hline
Fig. 5(d) & $T-m_0$ & Au-Au  & 19.6  & $0.185\pm0.003$  & $0.038\pm0.004$  & 0.15 \\
          &~        &~       & 27    & $0.186\pm0.003$  & $0.040\pm0.004$  & 0.14 \\
          &~        &~       & 39    & $0.190\pm0.003$  & $0.043\pm0.004$  & 0.19 \\
\hline
Fig. 5(e) & $T-m_0$ & Au-Au  & 62.4  & $0.179\pm0.003$  & $0.068\pm0.001$  & 0.01 \\
          &~        &~       & 130   & $0.179\pm0.003$  & $0.072\pm0.004$  & 0.03 \\
          &         & Au-Au  & 200   & $0.182\pm0.004$  & $0.073\pm0.004$  & 0.01 \\
\hline
Fig. 5(f) & $T-m_0$ & Pb-Pb  & 2760  & $0.214\pm0.003$  & $0.089\pm0.004$  & 0.45 \\
          &~        &~       & 5020  & $0.208\pm0.003$  & $0.114\pm0.003$  & 1.84 \\
          &~        & Xe-Xe  & 5440  & $0.216\pm0.003$  & $0.113\pm0.003$  & 1.23 \\
\hline
Fig. 6(a) & $\langle p_T\rangle-\overline m$ & Au-Au  & 2.7   & $0.114\pm0.004$  & $0.349\pm0.004$  & 0.99 \\
          &~                                 &        & 3.32  & $0.109\pm0.004$  & $0.376\pm0.005$  & 2.31 \\
          &~                                 &~       & 3.84  & $0.102\pm0.005$  & $0.387\pm0.005$  & 0.01 \\
\hline
Fig. 6(b) & $\langle p_T\rangle-\overline m$ & Au-Au  & 4.3   & $0.120\pm0.004$  & $0.389\pm0.005$  & 0.27 \\
          &~                                 &~       & 5.03  & $0.142\pm0.004$  & $0.379\pm0.005$  & 1.06 \\
          &~                                 & Pb-Pb  & 6.3   & $0.151\pm0.005$  & $0.372\pm0.004$  & 0.10 \\
\hline
Fig. 6(c) & $\langle p_T\rangle-\overline m$ & Au-Au  & 7.7   & $0.143\pm0.005$  & $0.403\pm0.005$  & 0.01 \\
          &~                                 &~       & 11.5  & $0.152\pm0.005$  & $0.398\pm0.007$  & 0.01 \\
          &~                                 &~       & 14.5  & $0.143\pm0.005$  & $0.408\pm0.006$  & 0.15 \\
\hline
Fig. 6(d) & $\langle p_T\rangle-\overline m$ & Au-Au  & 19.6  & $0.152\pm0.004$  & $0.407\pm0.005$  & 0.01 \\
          &~                                 &~       & 27    & $0.151\pm0.004$  & $0.412\pm0.006$  & 0.01 \\
          &~                                 &~       & 39    & $0.148\pm0.004$  & $0.422\pm0.006$  & 0.79 \\
\hline
Fig. 6(e) &$\langle p_T\rangle-\overline m$  & Au-Au  & 62.4  & $0.106\pm0.003$  & $0.466\pm0.006$  & 0.03 \\
          &~                                 &~       & 130   & $0.101\pm0.003$  & $0.472\pm0.008$  & 0.04 \\
          &~                                 &~       & 200   & $0.098\pm0.003$  & $0.484\pm0.008$  & 0.01 \\
\hline
Fig. 6(f) &$\langle p_T\rangle-\overline m$  & Pb-Pb  & 2760  & $0.090\pm0.002$  & $0.529\pm0.006$  & 0.01 \\
          &~                                 &~       & 5020  & $0.083\pm0.002$  & $0.539\pm0.008$  & 0.01 \\
          &~                                 & Xe-Xe  & 5440  & $0.090\pm0.002$  & $0.532\pm0.009$  & 0.01 \\
\hline
\end{tabular}%
\end{center}
\end{table*}

\begin{figure*}[!htb]
\centering
\includegraphics[width=13.0cm]{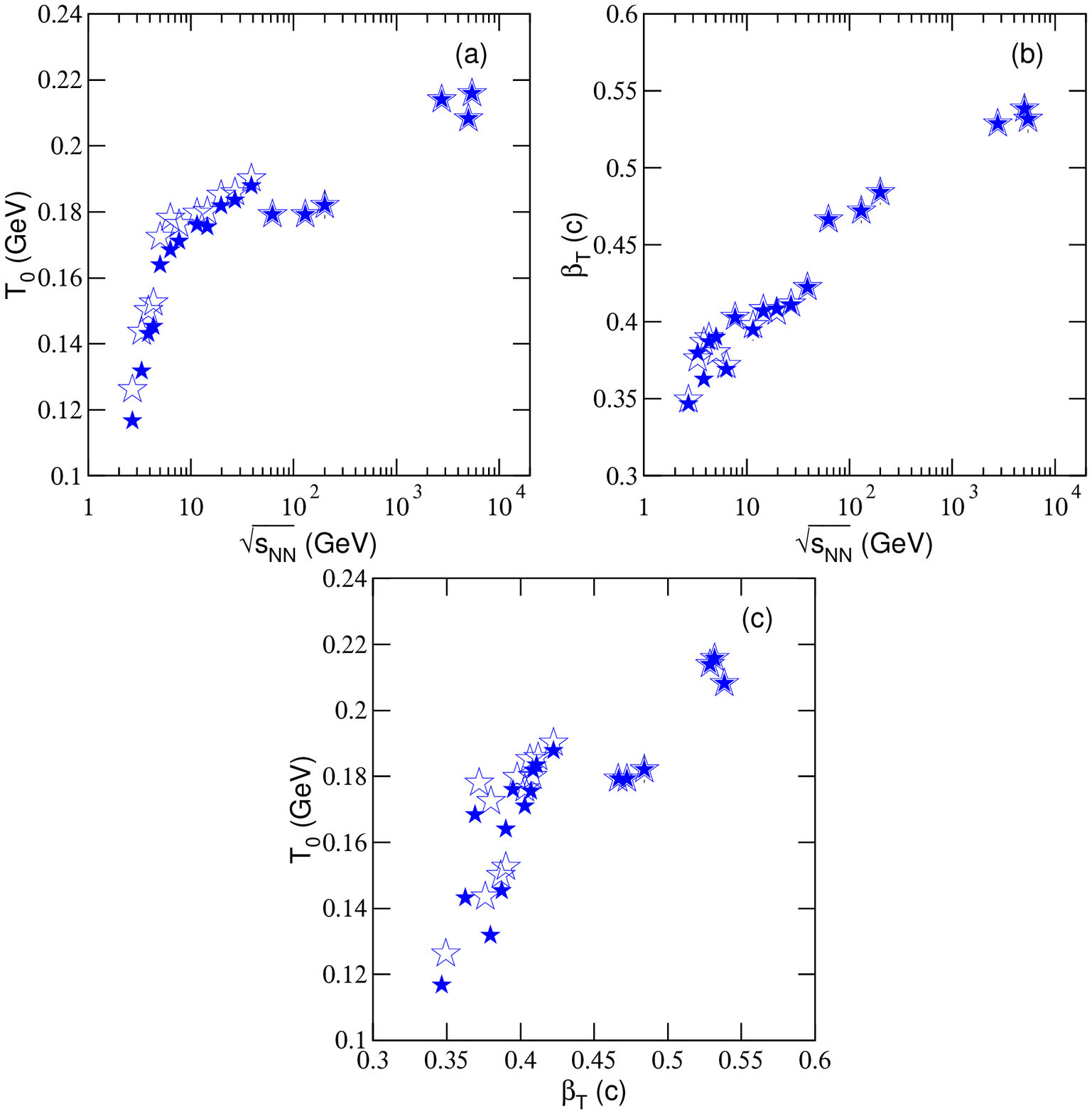}
\caption{Dependences of (a) $T_0$ on $\sqrt{s_{NN}}$, (b)
$\beta_T$ on $\sqrt{s_{NN}}$, and (c) $T_0$ on $\beta_T$. The
parameter values are obtained from Tables 3 and 4 which are from
the linear fittings in Figs. 5 and 6.} \label{F7}
\end{figure*}

The values of $a_0$ for the spectra of charged pions, kaons, and
protons at above 10 GeV are approximately around 0.75, 1, and 1.8,
respectively, which drop obviously for pions and kaons at lower
energy due to the hadronic phase. In addition, due to the
existence of participant protons in both the hadronic and QGP
phases, the energy dependence of $a_0$ for protons is not obvious.
Although it is hard to explain exactly the physical meaning of
$a_0$, we emphasize here that it shows the bending degree of the
spectrum in low-$p_T$ region~\cite{36b,36bb} and affects the
slopes in high-$p_T$ region due to the limitation of
normalization. A large bending degree means a large slope change.
In fact, $a_0$ is empirically related to the contributions of
strong decay from high-mass resonance and weak decay from heavy
flavor hadrons. This is because that $a_0$ affects mainly the
spectra in low-$p_T$ region which is just the main contribution
region of the two decays.

One can see that the values of $q$ and $a_0$ change drastically
with particle species. This is an evidence of mass-dependent
differential kinetic freeze-out scenario~\cite{26}. The massive
particles emit earlier than light particles in the system
evolution. The earlier emission is caused due to the fact that the
massive particles are left behind in the evolution process, but
not their quicker thermal and flow motion. In fact, the massive
particles have no quicker thermal and flow motion due to larger
mass. Instead, light particles have quicker thermal and flow
motion and longer evolution time. Finally, light particles reach
larger space at the stage of kinetic freeze-out.

The influence of $\mu_i$ on $q$ and $a_0$ is very small. Although
the prefactor $a_0$ can come from the Cooper-Frye term (and/or a
kind of saddlepoint integration) as discussed e.g. in
refs.~\cite{61d,68}, it is a fit parameter in this work. As an
average over pions, kaons, and protons, $\langle a_0\rangle$ is
nearly independent of $\sqrt{s_{NN}}$ at above 10 GeV. As
$\sqrt{s_{NN}}$ increasing from $\approx3$ to $\approx10$ GeV, the
increase of $\langle a_0\rangle$ shows different collision
mechanisms comparing with that at above 10 GeV. Our recent
work~\cite{66a} shows that the energy $\approx10$ GeV discussed
above is exactly 7.7 GeV.

\subsection{Derived parameters and their tendencies}

As we know, the effective temperature $T$ contains the
contributions of the thermal motions and flow effect~\cite{66aa}.
The thermal motion can be described by the kinetic freeze-out
temperature $T_0$, and the flow effect can be described by the
transverse flow velocity $\beta_T$. To obtain the values of $T_0$
and $\beta_T$, we analyze the values of $T$ presented in Tables 1
and 2, and calculate $\langle p_T\rangle$ and $\overline{m}$ based
on the values of parameters listed in Tables 1 and 2. In the
calculation performed from $p_T$ to $\langle p_T\rangle$ and
$\overline{m}$ by the Monte Carlo method, as in
refs.~\cite{24,25,26}, an isotropic assumption in the rest frame
of emission source is used.

Figures 5(a)--5(f) show the relationship of $T$ and $m_0$,
determined fitting AA collision systems by our model. Figures
6(a)--6(f) show the relationship of $\langle p_T\rangle$ and
$\overline m$, correspondingly. Different symbols represent the
values from central AA collisions at different $\sqrt{s_{NN}}$.
The symbols in Figs. 5(a)--5(f) represent the values of $T$ for
different $m_0$. The symbols in Figs. 6(a)--6(f) represent the
values of $\langle p_T \rangle$ for different $\overline m$.

We noted that, in Fig. 5(b), $T$ increases with the energy from
4.3 to 6.3 GeV for the emission of pions and not for protons,
while in the case 2.76--5.44 TeV in Fig. 5(f), $T$ increases for
the emission of protons and not for pions. This discrepancy
appears also when narrow energy ranges are fitted in experiments,
though $\langle p_T \rangle$ should rise for all particle species
as a function of $\sqrt{s_{NN}}$. We may explain this as the
fluctuations. It is expected that $T$ for emissions of both pions
and protons show the same or similar behavior with the energy in a
wide range.

It can be seen that the mentioned relationships show nearly linear
tendencies in most cases. The lines in Figs. 5 and 6 are the
results fitted by the least square method, where the solid and
dashed lines correspond to the results for $\mu_i=0$ and
$\mu_i=\mu_B/3$ respectively. The values of intercepts, slopes,
and $\chi^2$ are listed in Tables 3 and 4. One can see that, in
most cases, the mentioned relations are described by a linear
function. In particular, the intercepts in Figs. 5(a)--5(f) are
regarded as $T_0$, and the slopes in Figs. 6(a)--6(f) are regarded
as $\beta_T$, as what we discussed above in the alternative
method. Because different ``thermometers" are used, $T_0$
extracted from the intercept exceeds (is not in agreement with)
the phase transition temperature which is independently determined
by lattice QCD to be around 155 MeV. To compare the two
temperatures, we need a transform equation or relation which is
not available at present and we will discuss it later.

It is noted that, the above argument on $T_0$ and $\beta_T$ is
based usually on exact hydrodynamic calculations, as e.g. given in
refs.~\cite{17,61d,61a,61b,61c,61e}. But in these cases, usually
$T$ is extracted, and then some $T=T_0+m_0\langle u_t\rangle^2$
like correspondence is derived (where instead of $m_0$, also
energy or average energy could stand, depending on the
calculation). Here, as we know, $\langle u_t\rangle$ is related
but not equal to $\beta_T$, as discussed in the mentioned
literature. So, we give up to use $\langle u_t\rangle$ as
$\beta_T$ in this work.

We think that $T_0$ can be also obtained from $\langle
p_T\rangle$, and $\beta_T$ can be also obtained from $T$. However,
the relations between $T_0$ and $\langle p_T\rangle$, as well as
$\beta_T$ and $T$, are not clear. Generally, the parameters $T_0$
and $\beta_T$ are model dependent. In other models such as the
blast-wave model~\cite{17,18,19,20,21}, $T_0$ and $\beta_T$ can be
obtained conveniently. The two treatments show similar tendencies
of parameters on $\sqrt{s_{NN}}$ and event centrality, if we also
consider the flow effect in small system or peripheral AA
collisions~\cite{61f,61g} in the blast-wave model.

In order to more clearly see the tendencies of $T_0$ and
$\beta_T$, we show the dependences of $T_0$ on $\sqrt{s_{NN}}$,
$\beta_T$ on $\sqrt{s_{NN}}$, and $T_0$ on $\beta_T$ in Figs.
7(a)--7(c), respectively. One can see that the two parameters
increase quickly from $\approx3$ to $\approx10$ GeV and then
slowly at above 10 GeV with the increase of $\sqrt{s_{NN}}$ in
general. There is a plateau from near 10 GeV to 200 GeV. In
particular, $T_0$ increases with $\beta_T$ due to the fact that
both of them increase with $\sqrt{s_{NN}}$. These incremental
tendencies show that, in the stage of kinetic freeze-out, the
degrees of excitation and expansion of the system increase with
increasing $\sqrt{s_{NN}}$. These results are partly in agreement
with the blast-wave model which shows decreasing tendency for
$T_0$ and increasing tendency for $\beta_T$ with increasing
$\sqrt{s_{NN}}$ from the RHIC~\cite{3} to LHC~\cite{4} because
different partial $p_T$ ranges in the data are considered for
different particles, while this work uses the $p_T$ range as wide
as the data. The chemical potential shows obvious influence on
$T_0$ at the lower energies (below dozens of GeV). After
considering the chemical potential, the plateau in the excitation
function of $T_0$ becomes more obvious.

With the increase of $\sqrt{s_{NN}}$, the fact that the values of
$T_0$ and $\beta_T$ increase quickly from $\approx3$ to
$\approx10$ GeV and then slowly at above 10 GeV implies that there
are different collision mechanisms in the two energy ranges. In AA
collisions, if the baryon-dominated effect plays more important
role at below 10 GeV~\cite{50}, the meson-dominated effect should
play more important role at above 10 GeV. In the baryon-dominated
case, less energies are deposited in the system, and then the
system has low excitation degree and temperature. In the
meson-dominated case, the situation is opposite. Indeed,
$\approx10$ GeV is a particular energy which should be paid more
attention. It seems that the onset energy of deconfinement phase
transition from hadronic matter to QGP is possibly 10 GeV or
slightly lower (e.g. 7.7 GeV~\cite{66a}).

If we regard the plateau from near 10 to 200 GeV in the excitation
functions of $T_0$ and $\beta_T$ as a reflection of the formation
of QGP liquid drop, the quick increase of $T_0$ and $\beta_T$ at
the LHC is a reflection of higher temperature QGP liquid drop due
to larger energy deposition. At the LHC, the higher collision
energy should create larger energy density and blast wave, and
then higher $T_0$ and $\beta_T$. Although any temperature needs to
be bound by the phase transition on one side and free streaming on
the other side, larger energy deposition at the LHC may heat the
system to a higher temperature even the phase transition
temperatures at the LHC and RHIC are the same. Both the formed QGP
and hadronized products are also possible to be heated to higher
temperature.

Although we mentioned that the plateau apparent in $T_0$ versus
$\sqrt{s_{NN}}$ is possibly connected to the onset of
deconfinement, the temperature measured in this work is connected
only to $T_0$ which is usually much smaller than the quark-hadron
transition temperature. Because the collision process is very
complex, the $\sqrt{s_{NN}}$ dependence of $T_0$ reflects only
partial properties of the phase structure of a quark medium. To
make a determined conclusion, we may connect to the dynamics of
the hadron gas. This topic is beyond the focus of the present work
and will not be discussed further here.

We would like to point out that, in the last three paragraphs
mentioned above, the discussions on the excitation function of
$T_0$ presented in Fig. 7(a) are also suitable to the excitation
function of $T$ presented in Fig. 4(a), though the effect of flow
is not excluded from Fig. 4(a). Because the quality of fits is not
sufficient in a few cases, our main conclusion that the rise of
temperature below 10 GeV suggests that a deconfinement of hadronic
matter to QGP is weak. The information of phase transition
happened at higher temperatures and near the chemical freeze-out
may be reflected at the kinetic freeze-out of a hadronic system.
The plateau structure appeared in the excitation function $T_0$ is
expected to relate to the phase transition, though this relation
is not clear at present. Other works related to this issue are
needed to make a strong conclusion. In other words, to conclude
the onset of deconfinement just from the quality of some fits is a
loose interpretation. More investigations are needed and also
comparison with other findings. This issue is beyond the scope of
this analysis.

\subsection{Further discussion}

The model presented in the analysis can be regarded as a
``thermometer" to measure temperatures and other parameters at
different energies. Then, the related excitation functions can be
obtained and the differences from the transition around critical
point and other energies can be seen. Different models can be
regarded as different ``thermometers". The temperatures measured
by different ``thermometers" have to be unified so that one can
give a comparison. If we regard the phase transition temperature
determined by lattice QCD as the standard one, the values of $T_0$
obtained in this paper should be revised to fit the standard
temperature. However, this revision is not available for us at
present due to many uncertain factors. In fact, we try to focus on
the ``plateau" in the energy dependence of $T_0$, but not on the
$T_0$ values themselves.

In addition, the model assumes the contributions from two
participant partons in the framework of multisource thermal
model~\cite{32a}. In pp collisions, one can see the point of a
hard scattering between two partons and look at the high $p_T$
particle productions or other observations. However, even in pp
collisions there are underlying events, multiple-parton
interactions, etc. Further, the data used in this analysis are
from central AA collisions, where hundreds and thousands of
hadrons are produced. Although many partons take part in the
collisions, only a given two-parton process plays main role in the
production of a given set of particles. Many two-parton processes
exist in the collisions. Using a model inspired by two participant
partons is reasonable.

Of course, one may also expect that the production of many
particles can result from three or more partons. If necessary, we
may extend the picture of two participant partons to that of three
or multiple participant partons~\cite{32a} if we regard $p_T$ of
identified particle as the root sum square of the transverse
momenta of three or multiple participant partons. It is just that
the picture of two participant partons is enough for the
production of single particle in this analysis. Besides, we did
not try to distinguish between local thermalization of a
two-parton process. Instead, we regard the whole system as the
same temperature, though which is mass dependent.

The present work is different from the quark coalescence
model~\cite{68,63,64,65,66,67}, though both the models have used
the thermalization and statistics. In particular, the quark
coalescence model describes classically mesonic prehadrons as
quark-antiquark clusters, and baryonic ones composed from three
quarks. The present work describes both mesons and baryons as the
products of two participant partons which are regarded as two
energy sources.

The assumption of two participant partons discussed in the present
work does not mean that the particles considered directly stem
from two initial partons of the incoming nuclei. In fact, we
assume the two participant partons from the violent collision
system in which there is rescattering, recombination, or
coalescence. The two participant partons are only regarded as two
energy sources to produce a considered particle, whether it is a
meson or a baryon, or even a lepton~\cite{36b,36bb}. The present
work treats uniformly the production of final-state particles from
the viewpoint of participant energy sources, but not the quark
composition of the considered particles~\cite{68,63,64,65,66,67}.

In the two-component distribution [Eq. (10)], the first component
contributed by the soft excitation process is from the sea quarks.
The second component contributed by the hard scattering process is
from the valence quarks. This explanation is different from the
Werner's picture on core-corona separation~\cite{69,70,71,72} in
which core and corona are simply defined by the density of partons
in a particular area of phase or coordinate-space and they
distinguish between thermal and non-thermal particle production.
This could also be a two-component fit based on the Tsallis
function, but its relation to the two-parton dynamics pushed here
is not clear. Anyhow, it is possible that the two processes can be
described by a uniform method~\cite{36b,36bb}, though different
functions and algorithms are used.

Although there were many papers in the past that have studied the
identified particle spectra in high energy collisions, both
experimentally and phenomenologically, this work shows a new way
to systemize the experimental data in AA collisions over a wide
energy range from 2.7 GeV to 5.44 TeV at the parton level. We
emphasize that, in this work, we have analyzed the particle $p_T$
as the root sum square of transverse momenta \(p_{t1}\) and
\(p_{t2}\) of two participant partons. That is, the relation of
\(p_T=\sqrt{p_{t1}^2+p_{t2}^2}\) is used. While, in our recent
work~\cite{36b,36bb}, the relation of \(p_T=p_{t1}+p_{t2}\) is
used, which is considered from energy relation at mid-$y$ for
massless particle. The scenarios used in this work and our recent
work are different. Based on our analyses, it is hard to judge
which scenario is more reasonable.

Through the analysis of the data, we have obtained the excitation
functions of some quantities, such as $T$ and its weighted average
$\langle T\rangle$, $T_0$ and its weighted average $\langle
T_0\rangle$, $\beta_T$ and its weighted average $\langle
\beta_T\rangle$, $q$ and its weighted average $\langle q\rangle$,
as well as $a_0$ and its weighted average $\langle a_0\rangle$.
These excitation functions all show some specific laws as
$\sqrt{s_{NN}}$ increases. Although the conclusion on ``onset of
deconfinement" or QCD phase transition is indicated around 10 GeV
or below is possibly over-interpreting the data and only using the
blast-wave or Tsallis-like model is clearly not enough, the sudden
change in the slope in the excitation function of $T_0$ is worthy
of attention.
\\

\section{Summary and conclusion}

We summarize here our main observations and conclusions.

(a) The transverse momentum (mass) spectra of charged pions,
kaons, and protons produced at mid-rapidity in central AA (Au-Au,
Pb-Pb, and Xe-Xe) collisions over an energy range from 2.7 GeV to
5.44 TeV have been analyzed in this work. The experimental data
measured by several collaborations are fitted satisfactorily in
the framework of multisource thermal model in which the transverse
momentum of identified particle is regarded as the root sum square
of transverse momenta of two participant partons, where the latter
obeys the revised Tsallis-like function. This treatment for the
spectra of transverse momenta is novel and successful. The
excitation functions of parameters such as the effective
temperature, entropy index, revised index, kinetic freeze-out
temperature, and transverse flow velocity are obtained. The
chemical potential has obvious influence on the excitation
function of kinetic freeze-out temperature at lower energy.

(b) With increasing collision energy, the entropy index increases
slowly, and the revised index increases quickly and then changes
slowly except for a large increase at the LHC. With increasing the
particle mass, the entropy index decreases and the revised index
increases obviously. The collision system discussed in this work
stays approximately in the equilibrium state, and some functions
based on the assumption of equilibrium can be used. The system is
closer to the equilibrium state when it emits protons at lower
energy, comparing with pions and kaons at higher energy. The
revised index describes the bending degrees of the spectra in very
low transverse momentum region. Its values for the spectra of
charged pions, kaons, and protons are approximately around 0.75,
1, and 1.8, respectively, at above 10 GeV and drop obviously at
below 10 GeV.

(c) With increasing collision energy, the effective temperature
increases clearly and monotonously, and the kinetic freeze-out
temperature and transverse flow velocity increase quickly from
$\approx3$ to $\approx10$ GeV and then slowly at above 10 GeV.
There is a plateau from near 10 GeV to 200 GeV in the excitation
functions of the latter pair. The onset energy of deconfinement
phase transition from hadronic matter to QGP is connected to the
special changes of excitation function of kinetic freeze-out
temperature and possibly 10 GeV or slightly lower. If the plateau
at the RHIC is regarded as a reflection of the formation of QGP
liquid drop, the following quick increase of the excitation
functions at the LHC is a reflection of high temperature QGP
liquid drop due to larger energy deposition. At kinetic
freeze-out, the temperature and expansion velocity of the system
increase with increasing the energy from the RHIC to LHC.
\\
\\
\\
{\bf Author Contributions:} All authors contributed equally to
this work. All authors have read and agreed to the published
version of the manuscript.
\\
\\
{\bf Funding:} The work of L.L.L. and F.H.L. was supported by the
National Natural Science Foundation of China under Grant Nos.
12047571, 11575103 and 11947418, the Scientific and Technological
Innovation Programs of Higher Education Institutions in Shanxi
(STIP) under Grant No. 201802017, the Shanxi Provincial Natural
Science Foundation under Grant No. 201901D111043, and the Fund for
Shanxi ``1331 Project" Key Subjects Construction. The work of
K.K.O. was supported by the Ministry of Innovative Development of
Uzbekistan within the fundamental project on analysis of open data
on heavy-ion collisions at RHIC and LHC.
\\
\\
{\bf Institutional Review Board Statement:} Not applicable.
\\
\\
{\bf Informed Consent Statement:} Not applicable.
\\
\\
{\bf Data Availability Statement:} The data used to support the
findings of this study are included within the article and are
cited at relevant places within the text as references.
\\
\\
{\bf Conflicts of Interest:} The authors declare no conflict of
interest. The funding agencies have no role in the design of the
study; in the collection, analysis, or interpretation of the data;
in the writing of the manuscript, or in the decision to publish
the results.
\\
\\
{\bf Compliance with Ethical Standards:} The authors declare that
they are in compliance with ethical standards regarding the
content of this paper.
\\
\\


\end{multicols}
\end{document}